\documentclass[a4paper,12pt]{article}

 \pdfoutput=1

\usepackage[a4paper,text={16.8cm,22.8cm}]{geometry}
\usepackage{amsmath,amssymb,bm,graphicx,color}
\usepackage{extarrows}
\usepackage[utf8]{inputenc}
\usepackage{amsthm}
\usepackage{slashed}
\usepackage{tikz}
\usepackage{rotating}
\usepackage{hyperref}
\usepackage{array}
\usepackage{color}
\usepackage{multirow}
\usepackage{bbold}
\usepackage{cite}
\usepackage{pifont}
\usepackage{tabularx}

\allowdisplaybreaks 
\renewcommand{\arraystretch}{1.2}

\begin{document}

\begin{titlepage}

\vspace*{-1.0cm}
\begin{flushright} 
P3H-20-021\\
TTP20-022
\end{flushright}

\vspace{1.2cm}
\begin{center}
\Large\bf
\boldmath
Reconstructing Effective Lagrangians Embedding Residual Family Symmetries
\unboldmath
\end{center}
\vspace{0.2cm}
\begin{center}
{\large{Jordan Bernigaud$^{a,\,b}$, Ivo de Medeiros Varzielas$^c$, and Jim Talbert$^d$}}\\
\vspace{1.0cm}
{\sl 
${}^a$\,Institute for Nuclear Physics (IKP), Karlsruhe Institute of Technology, Hermann-von-Helmholtz-Platz 1, D-76344 Eggenstein-Leopoldshafen, Germany,\\[0.3cm]
${}^b$\,Institute for Theoretical Particle Physics (TTP), Karlsruhe Institute of Technology, Engesserstrasse 7, D-76128 Karlsruhe, Germany\\[0.3cm]
${}^c$\,CFTP, Departamento de F\'{i}sica, Instituto Superior T\'{e}cnico, Universidade de Lisboa, Avenida Rovisco Pais 1, 1049 Lisboa, Portugal\\[0.3cm]
${}^d$\,Niels Bohr International Academy, Niels Bohr Institute, University of Copenhagen, Blegdamsvej 17, DK-2100 Copenhagen, Denmark}\\[0.5cm]
{\bf{E-mail}}:  jordan.bernigaud@kit.edu, ivo.de@udo.edu, ronald.talbert@nbi.ku.dk
\end{center}

\vspace{0.5cm}
\begin{abstract}
\vspace{0.2cm}
\noindent 
We consider effective Lagrangians which, after electroweak- and family-symmetry breaking, yield fermionic mass matrices and/or other flavoured couplings exhibiting residual family symmetries (RFS).  Thinking from the bottom up, these RFS intimately link ultraviolet (UV) Beyond-the-Standard Model (BSM) physics to infrared flavour phenomenology without direct reference to any (potentially unfalsifiable) UV dynamics.  While this discussion is typically performed at the level of RFS group generators and the UV flavour groups they can close, we now also focus on the RFS-implied shape of the low-energy mass/coupling matrices. 
We then show how this information can be used to algorithmically guide the reconstruction of an effective Lagrangian, thereby forming top-down models realizing the typical bottom-up phenomenological conclusions.  As a first application we take results from scans of finite groups capable of controlling (through their RFS) CKM or PMNS mixing within the SM alone.  We then extend this to recently studied scenarios where RFS also control special patterns of leptoquark couplings, thus providing proof-in-principle completions for such `Simplified Models of Flavourful Leptoquarks.' 
\end{abstract}
\vfil

\end{titlepage}


\tableofcontents
\noindent \makebox[\linewidth]{\rule{16.8cm}{.4pt}}


\section{Introduction}
\label{sec:INTRO}

The unexplained 20-22 free and physical parameters associated to the masses, mixings, and CP-violating phases of the Standard Model's (SM) flavour sector (the so-called \emph{Flavour Puzzle}) represents an open challenge for theoretical constructions Beyond the SM (BSM).  While these parameters are technically natural, their appearance in the quark sector is associated to an explicit breaking of the U(3)$^5$ global flavour symmetry otherwise present in the SM Lagrangian \cite{Gerard:1982mm,Chivukula:1987py}, while the observation of neutrino masses is already a definitive new physics phenomenon.  Furthermore, the actual values of fermionic masses and mixings exhibit tantalizing hierarchies, including dramatically different patterns between quark and lepton sectors.  These observations beg for a dynamical origin for flavour, and countless BSM models based on family symmetries have been devised to that end, with some even attempting explanations for the presence of the otherwise arbitrary flavour index ($i = 1,2,3$) in the first place.

However, the model space is underdetermined --- multiple models based on different symmetries can predict the same phenomenology, and often models based on the same family symmetry can yield different infrared (IR) predictions when (unfalsifiable) tweaks to ultraviolet (UV) Lagrangian parameters are made.  Indeed,  it may be impossible to determine a true theory of flavour in the absence of any convincing observation of new physics that distinguishes SM fermion generations, especially since reliable experimental constraints already exist for all but the leptonic (Dirac) CP-violating phase, absolute neutrino masses, and additional parameters depending on whether neutrinos are Majorana particles.  Therefore most model predictions should actually be considered `post-dictions.' 

One might then pursue a formalism for describing BSM flavour in more model-independent ways, focusing only on connecting patterns of family-symmetry breaking (which can themselves be generically motivated, perhaps in stringy theories --- see \cite{Baur:2019kwi,Baur:2019iai}, e.g.) to the relevant IR phenomenology, and not on unfalsifiable Lagrangians based on new heavy states or dynamics that may be associated to that symmetry breaking.  Residual Family Symmetries (RFS) provide just such a formalism, as they promote accidental Abelian symmetries of the SM mass sector to the residual subgroups of a UV flavour symmetry $\mathcal{G}_F$.  In bases where the physical mixing parameters appear in the SM Yukawa Lagrangian (any basis other than the mass-eigenstate basis), one then notes that the Abelian generators associated to RFS are themselves functions of the physical mixing parameters.  Closing flavour groups with these generators then provides the desired, model-independent link between UV symmetries and IR mixing phenomenology, and multiple analytic and computational studies have been performed to uncover viable $\mathcal{G}_F$ \cite{Lam:2007qc,Ge:2011ih,Ge:2011qn,Grimus:2011fk,deAdelhartToorop:2011re,Hernandez:2012ra,Lam:2012ga,Holthausen:2012wt,Holthausen:2013vba,King:2013vna,Lavoura:2014kwa,Fonseca:2014koa,Hu:2014kca,Joshipura:2014pqa,Joshipura:2014qaa,Yao:2015dwa,King:2016pgv,Yao:2016zev,Lu:2016jit,Li:2017abz,Lu:2018oxc,Hagedorn:2018gpw,Lu:2019gqp,Talbert:2014bda, Varzielas:2016zuo,Blum:2007jz}.

Of course, if a particular `simplified model' (or class of simplified models) based on the RFS formalism is singled out due to new measurements in the flavour sector, a more complete description of the physics will be desired. In this paper we provide a method to (re)construct effective Lagrangians that recover the symmetry breaking distilled in RFS scenarios.  That is, we show how to construct a top-down model from a bottom-up phenomenological observation/conclusion.  We do so by focusing on the intimate link between RFS generators and the implied shape of an RFS-invariant mass/coupling matrix.  After all, RFS are symmetries of mass matrices and not the full SM Lagrangian and so, up to possible ambiguities associated to the group-theoretical properties of RFS generators, a specific symmetry-breaking pattern from the UV $\mathcal{G}_F$ to a given RFS implies a specific IR mass/coupling matrix.  This shape then hints at relevant multiplet charge assignments under the parent $\mathcal{G}_F$, which when combined with RFS-implied vacuum expectation values (VEV) for family-symmetry breaking scalar flavons, can be used to algorithmically construct an effective Lagrangian.  We first apply this method to models addressing SM mixing structures alone, i.e. the $U_{PMNS}$ or $U_{CKM}$ matrices, and then also to a class of `Simplified Models of Flavourful Leptoquarks' developed in \cite{deMedeirosVarzielas:2019lgb,Bernigaud:2019bfy}.  These models include a new Yukawa-like coupling between the leptoquark and SM quark and lepton doublets which is, in addition to the SM mixing, controlled by RFS.  They therefore generate rich, flavour-dependent phenomenology at the Large Hadron Collier (LHC) and other precision experiments which can be used to probe their predictions.

The paper develops as follows:  in Section \ref{sec:THEORY} we give a pedagogical review of the RFS formalism, making explicit the intimate connection between RFS generators and implied mass shapes, while also describing bottom-up techniques to close UV flavour groups.
We then discuss how to take those results and build an effective UV Lagrangian.  In Section \ref{sec:SM} we apply this recipe to models reproducing $U_{PMNS}$ or $U_{CKM}$ before moving to leptoquark applications in Section \ref{sec:LEPTOQUARKS}.  We conclude in Section \ref{sec:CONCLUDE}, give relevant information for the finite groups we employ in Appendix \ref{sec:APP}, and also give further details on our core RFS-preserving flavon condition in Appendix \ref{sec:APPB}.

\section{RFS:  Bottom-Up Formalism for Top-Down Models}
\label{sec:THEORY}
The core assumption of the RFS paradigm is that a parent flavour symmetry $\mathcal{G_F}$ is broken in such a way that, after subsequent EWSB, RFS mediated by subgroups of $\mathcal{G_F}$ are preserved in some or all of the SM mass matrices, or indeed any other term controlled by the original flavour symmetry.  For example, a natural symmetry breaking pattern through intermediate groups controlling the lepton and quark sectors of the SM is schematically illustrated by
\begin{equation}
\label{eq:GF}
\mathcal{G_{F}}  \rightarrow \begin{cases}
				\mathcal{G_{L}}   \rightarrow \begin{cases} 
										\mathcal{G_{\nu}}
										\\
										\mathcal{G_{\text{l}}}
										\end{cases} \\
				\mathcal{G_{Q}} \rightarrow \begin{cases}
										\mathcal{G_{\text{u}}}
										\\
										\mathcal{G_{\text{d}}} \,.
										\end{cases}
				\end{cases}
\end{equation}
Of course other patterns beyond \eqref{eq:GF}, perhaps without the intermediate $\mathcal{G_{L,Q}}$ or which only address either the quark or lepton sector individually, are also conceivable.  $\mathcal{G_{F,L,Q}}$ can in principle be Abelian or non-Abelian, continuous or discrete, although for the remainder of this paper we will assume that $\mathcal{G_{F,L,Q}}$ are non-Abelian, such that irreducible multiplets of dimension greater than one can be arranged in flavour space.  Furthermore, we will only work with non-Abelian discrete symmetries (NADS) when constructing explicit models in Sections \ref{sec:SM}-\ref{sec:LEPTOQUARKS}, although our general approach and analysis in this section is equally applicable to non-Abelian continuous flavour groups as well.  Also, the RFS $\mathcal{G_{\text{a}}}$ with $a \in \lbrace u, d, l, \nu \rbrace$ must be Abelian, and when considering they reconstruct NADS we have in particular  Abelian cyclic groups of order $n$,
\begin{equation}
\label{eq:cyclic}
\mathcal{G}_{a} \cong \mathbb{Z}_{n_a}\,\,,
\end{equation}
when $\mathcal{G_{F,L,Q}}$ are themselves discrete.  Discrete product groups of the form $\mathcal{G}_{a} \sim \mathbb{Z}_{n_a,1} \times \mathbb{Z}_{n_a,2} \times \text{...}$ are also possible.  Finally, we note that the complete flavour symmetry present in the effective Lagrangians to be considered in the upcoming sections is actually $\mathcal{G_{F} } \times  \mathcal{G_{\text{shape}}}$, where $\mathcal{G_{\text{shape}}}$ is an Abelian \emph{shaping symmetry} that forbids unwanted scalar interactions.  However, as will become clear, unlike $\mathcal{G}_{a}$ we do not need to explicitly specify $\mathcal{G_{\text{shape}}}$ ab initio, as our bottom-up RFS approach depends only on the fact that $\mathcal{G_{\text{shape}}}$ can be found to exist \emph{after} realizing all of the desired phenomenology.
\subsection{The Infrared Lagrangian} 
\label{sec:REV}
To review how the RFS chain in \eqref{eq:GF} is naturally motivated, we follow prior discussions (see e.g. \cite{Lam:2007qc,Hernandez:2012ra}) and first write down the SM Yukawa Lagrangian, after EWSB, in the mass-eigenstate basis:
\begin{align}
\label{eq:SMfermionmass}
\mathcal{L}^{SM}_{mass} \,\,\, \supset \,\,\,&\frac{1}{2} \bar{\nu}^{c}_{L}\, m_{\nu}\, \nu_{L} + \bar{E}_{R}\, m_{l}\, l_{L} + \bar{d}_{R}\, m_{d}\, d_{L} + \bar{u}_{R}\, m_{u} \,u_{L} +\, \text{h.c.}\,,
\end{align}
where $m_a$ are diagonal matrices of mass eigenvalues and where we have assumed a Majorana neutrino mass term to illustrate our point, although our RFS approach applies straightforwardly to a Dirac mass $\propto \bar{\nu}_R\, m_\nu\, \nu_L$ as well.  From \eqref{eq:SMfermionmass} we observe that the Lagrangian is invariant under Abelian transformations on the fermion fields:\footnote{Other basis choices can be made for the Klein generators $T_{\nu_i}$.}
\begin{align}
\nonumber
\nu_{L} &\rightarrow T_{\nu_{i}}\nu_{L}, \,\,\,\,\,&&\text{with} \,\,\,\,\,&&T_{\nu 1} = diag \left(1,-1,-1 \right) &&\text{and}\,\,\,\,\, T_{\nu 2} = diag \left(-1,1,-1 \right),\\
\label{eq:SMRFS}
f &\rightarrow T_{f} f, \,\,\,\,\,&&\text{with} \,\,\,\,\,&&T_{f} = diag \left(e^{i \alpha_{f}},e^{i \beta_{f}},e^{i \gamma_{f}} \right)\,\,\,\,\,&&\text{for}\,\,\,\,\, f \in \lbrace E_{R}, l_{L}, d_{R}, d_{L}, u_{R}, u_{L} \rbrace.
\end{align}
Hence we promote these accidental symmetries to RFS, and note that, for the case of Majorana neutrinos, the (maximal) RFS generated by $T_{\nu_i}$ is a Klein four-group \cite{Lam:2007qc}, 
\begin{equation}
\label{eq:Klein}
\mathcal{G_\nu} \cong \mathbb{Z}_2^\nu \times \mathbb{Z}_2^\nu \,.
\end{equation}
Similarly, the generators $T_f$ simply represent re-phasing freedoms of the three fermion generations in each family's Dirac mass term, 
\begin{equation}
\label{eq:U13}
\mathcal{G_\text{f}} \cong U(1)^3\,,
\end{equation}
and of course a Dirac neutrino mass term would be generically invariant under \eqref{eq:U13} instead of \eqref{eq:Klein}.  When \eqref{eq:cyclic} is realized the otherwise continuous phases in $T_f$ are quantized as
\begin{equation}
\label{eq:discretizephases}
\lbrace \alpha, \beta, \gamma \rbrace_{f} \overset{!}{=} \frac{2 \pi}{m} \lbrace a, b,  c  \rbrace_{f}\,.
\end{equation}
Finally, from \eqref{eq:SMfermionmass} one also finds that $T_{f_L} \overset{!}{=} T_{f_R}$ for the terms to be invariant.  Of course,  left- and right-chiral fermions can be charged differently  in the complete flavour theory invariant under $\mathcal{G_{F,L,Q}}$.  

However, \eqref{eq:SMfermionmass}-\eqref{eq:SMRFS} tell us nothing about the physical predictions associated to the family symmetry breaking in \eqref{eq:GF}.  It is only when we rotate to a basis where the Yukawa terms contain information about fermionic mixing that the RFS is useful as a bottom-up tool.  Take the standard `flavour basis' of the SM, where charged-current (CC) interactions are diagonal, as an example.  Here \eqref{eq:SMfermionmass} is transformed to
\begin{align}
\label{eq:SMfermionflav}
\mathcal{L}^{SM}_{flav} \,\,\, \supset \,\,\,&\frac{1}{2} \bar{\nu}^{c}_{L}\underbrace{U^{\star}_{\nu} m_{\nu} U^{\dagger}_{\nu}}_{m_{\nu U}} \nu_{L} + \bar{E}_{R}\underbrace{U_{E} m_{l}U_{l}^{\dagger}}_{m_{lU}} l_{L} + \bar{d}_{R}\underbrace{U_{D} m_{d} U_{d}^{\dagger}}_{m_{dU}} d_{L} + \bar{u}_{R}\underbrace{U_{U}  m_{u} U_{u}^{\dagger}}_{m_{u U}} u_{L} +\, \text{h.c.} \,,
\end{align}
where the left-handed unitary matrices $U$ have physical effects in the CC through the presence of the CKM and PMNS overlap matrices:
\begin{equation}
\label{eq:CKMPMNS}
U_{CKM} \equiv U_u^\dagger \, U_d, \,\,\,\,\,\,\,\,\,\,U_{PMNS} \equiv U_l^\dagger \, U_\nu \,.
\end{equation}
Both $U_{CKM,PMNS}$ are $3 \times 3$ matrices in flavour space and are parameterized by three mixing angles $\theta_{12,23,13}^{q,l}$ and one Dirac CP-violating phase $\delta^{q,l}$.  If neutrinos are Majorana particles, the PMNS also encodes two additional phases $\alpha_{1,2}$.  Hence it is clear that the redefined mass matrices $m_{aU}$ (where $a$ denotes all fermions) in \eqref{eq:SMfermionflav} are themselves $3 \times 3$ matrices in flavour space, and are of course related to the SM Yukawa couplings $Y_a$ through the Higgs VEV $v$, 
\begin{equation}
\label{eq:massYuk}
m_{aU} \equiv \frac{v}{\sqrt{2}} Y_a\,.
\end{equation}
Obviously \eqref{eq:massYuk} does not hold for the Majorana neutrino mass term written explicitly above.

Let us now examine the RFS of \eqref{eq:SMfermionflav}.  Here one observes that the Lagrangian is invariant under transformations of the form
\begin{equation}
\label{eq:flavbasisTa}
a \rightarrow T_{aU} \, a, \,\,\,\,\,\text{with} \,\,\,\,\, T_{aU} = U_a\,T_a\,U_a^\dagger \,,
\end{equation}
as opposed to those of \eqref{eq:SMRFS}.  Indeed, the RFS generator $T_{aU}$ now knows about the physical mixing matrices $U_a$, which means that any parent group $\mathcal{G_{F,L,Q}}$ with subgroup $\mathcal{G_\text{a}}$ generated by $T_{aU}$ can be connected to a physical mixing prediction embedded in $U_a$.  In this way the RFS intimately links the IR phenomenology to the UV symmetry without reference to any of the dynamics associated to realizing \eqref{eq:GF}.  RFS therefore provide a powerful, bottom-up means of understanding observed patterns of flavour mixing in a rather model-independent way, as the only assumption made thus far is that the accidental flavour symmetries of the SM mass sector encoded in \eqref{eq:SMRFS} are in fact the global RFS of a complete flavour theory broken as in \eqref{eq:GF} or its analogues.\footnote{The RFS formalism therefore characterizes a class of flavour models, sometimes referred to as `direct' or `semi-direct' in the literature \cite{King:2013eh}, but it is of course not entirely generic.  It is plausible that \eqref{eq:SMRFS} instead represent truly accidental symmetries, and that $T_a$ do not generate the subgroups of $\mathcal{G_{F,L,Q}}$.  These types of models are sometimes called `indirect' --- see \cite{deMedeirosVarzielas:2017sdv} for a successful GUT-inspired example.}  In this way the RFS formalism defines a set of `simplified flavour models,' which can easily be extended to BSM constructions as well --- see \cite{deMedeirosVarzielas:2019dyu} for a recent application of RFS to the Yukawa sector of multi-Higgs doublet models, where they were shown to be capable of controlling dangerous flavour-changing neutral currents alongside of fermionic mixing, and \cite{deMedeirosVarzielas:2019lgb,Bernigaud:2019bfy} where it is demonstrated that they can also structure the flavour patterns of leptoquark couplings (we also address some of these models in Section \ref{sec:LEPTOQUARKS} below).

However, \eqref{eq:SMfermionflav} is but one of an infinite number of bases that the Yukawa sector can be written in.  In the presence of  BSM couplings that introduce new (physical) mixings, one may want to work in a different one in order to preserve diagonal CC (see the `leptoflavour basis' discussed in Section \ref{sec:LEPTOQUARKS}).  Or one may be motivated to change basis due to the ease of use of certain (basis-dependent) group product rules.
Regardless, the trend as regards the associated RFS symmetry transformation is trivially clear; a rotation on a mass-eigenstate field $a$ with unitary matrix $V^\dagger_a$ equivalently implies a basis-change on the corresponding RFS mass-basis generator $T_a$ through the same matrix:
\begin{equation}
\label{eq:correspondence}
a \rightarrow V^\dagger_a \, a \,\,\, \Longleftrightarrow \,\,\, T_a \rightarrow V_a \, T_a \, V^{\dagger}_a \,.
\end{equation}
The statement holds vice versa as well, since otherwise $T_a$ would no longer be an RFS generator, as it would not leave the associated mass/Yukawa term invariant.  In Section \ref{sec:MODEL} we will apply the logically equivalent statement to \eqref{eq:correspondence} to study the RFS-invariant mass matrices themselves, in an effort to guide the reconstruction of an effective Lagrangian with manifest $\mathcal{G_{F,L,Q}}$.

Before doing so,  it is important to address a couple of subtleties in the approach, for clarity.  First, the RFS are not symmetries of the full IR Lagrangian.   CC interactions do not respect them without additional assumptions relating RFS within the quark and/or lepton sector.  This is realized naturally in most flavour models, however, since the breaking of $\mathcal{G_{F,L,Q}}$ is typically only communicated to the Yukawa sector, perhaps through scalar flavons developing VEV.  This is the approach we will take in what follows, although it is worth noting that family-symmetry breaking can also occur through other mechanisms, e.g. orbifold compactifications.

And secondly, we recall that a bottom-up RFS analysis alone cannot recover the exact mixing prediction associated to the model sketched by \eqref{eq:GF} unless $\mathcal{G_\text{a}}$ distinguishes all three fermion generations, i.e. the associated RFS generator(s) $T_a$ needs to have three distinct eigenvalues (or multiple $T_{a_i}$ need to be present when $T_a$ has fewer than three distinct eigenvalues).  This point becomes clear in the following flavour-basis equality: 
\begin{equation}
\label{eq:degenEV}
T_{aU} = U_{a}\,T^{ii=jj}_{a} \, U_{a}^{\dagger} = U_{a}\,R_{a}^{ij} \,T^{ii=jj}_{a} \,R_{a}^{ji\star}\, U_{a}^{\dagger}, \,\,\,\,\,\text{with} \,\,\,\,\, R^{ij} \equiv 
\left(
\begin{array}{cc}
\cos \theta_{ij} & \sin \theta_{ij} \,e^{-i\delta_{ij}} \\
-\sin \theta_{ij} \,e^{i\delta_{ij}} & \cos \theta_{ij}
\end{array}
\right). 
\end{equation}
That is, the RFS generator cannot distinguish between the mixing matrix $U_a$ and $U_a \cdot R_a$, with the latter having free parameters in the degenerate $(i,j)$ sector of $T_a$.  In complete models these free parameters can either be fit to data or quantized as a result of other mechanisms, like further auxiliary or accidental symmetries of the Lagrangian.  We will discuss the top-down implications of \eqref{eq:degenEV} in upcoming sections. 

\subsection{Closing Ultraviolet Flavour Groups} 
\label{sec:SCAN}
Given \eqref{eq:correspondence}, one must then have a procedure for recovering the associated parent groups $\mathcal{G_{F,L,Q}}$, as the Abelian $\mathcal{G_{\text{a}}}$ alone are insufficient to model patterns of physical mixing. 
Many groups have attempted this by performing either analytic or computational studies of the classes of $\mathcal{G_{F,L,Q}}$ that can break to desired subgroups $\mathcal{G_\text{a}}$, given that specific (phenomenologically viable) shapes for $U_{a}$ must be achieved in a realistic model.  On the other hand, the {\tt{GAP}} language for computational finite algebra \cite{Sch97,GAP4} has been indispensable when searching for NADS with automated techniques, as it has a large library of small groups catalogued along with vast amounts of associated group-theoretical information (conjugacy classes, order, irreducible representations, etc.).  

In what follows we will use the bottom-up approach to `reconstructing' NADS first discussed in \cite{Talbert:2014bda,Varzielas:2016zuo}, but recently applied to a class of BSM leptoquark models in \cite{deMedeirosVarzielas:2019lgb,Bernigaud:2019bfy}.  Here one assumes that the RFS generators form the complete generating set for $\mathcal{G_{F,L,Q}}$, such that the latter are recovered upon using {\tt{GAP}} to close all elements of the former:
\begin{align}
\label{eq:GFall}
\mathcal{G_F} &\cong \lbrace \hat{T}_{d}, \hat{T}_{l},\hat{T}_{u}, \hat{T}_{\nu} \rbrace\,, \\
\label{eq:GQGL}
\mathcal{G_F} &\cong \underbrace{\lbrace \hat{T}_{d},\hat{T}_{u} \rbrace}_{\mathcal{G_Q}} \times \underbrace{\lbrace \hat{T}_{l}, \hat{T}_{\nu}\rbrace}_{\mathcal{G_L}}\,, \\
\label{eq:GL}
\mathcal{G_L} &\cong \lbrace \hat{T}_{l}, \hat{T}_{\nu} \rbrace \,,\\
\label{eq:GQ}
\mathcal{G_Q} &\cong \lbrace \hat{T}_{u}, \hat{T}_{d} \rbrace \,,
\end{align}
where \eqref{eq:GFall} reconstructs a parent group generated by all family sectors, \eqref{eq:GQGL} forms a direct product parent group of lepton and quark symmetries, and \eqref{eq:GL}-\eqref{eq:GQ} assume that the NADS only controls either lepton or quark mixing, but not both.  Other scenarios could also be envisaged, e.g. one where $\mathcal{G_{L,Q}}$ are formed as in \eqref{eq:GL}-\eqref{eq:GQ}, but where $\mathcal{G_F}$ is not their direct product group as in \eqref{eq:GQGL}, but instead any larger group containing $\mathcal{G_{L,Q}}$.  Regardless, the hatted ($\hat{T}$) notation in \eqref{eq:GFall}-\eqref{eq:GQ} simply indicates any basis where the generators know about physical mixing parameters,
\begin{equation}
\hat{T}_a \equiv \hat{T}_a \left( \theta^a_{ij}, \delta^a_{ij}, ... \right)\,.
\end{equation}
Of course, when searching for NADS in the bottom-up approach one must also apply a discretization scheme to all of these mixing parameters, and details on this procedure and other cuts made regarding group order, etc., can be found on case-by-case bases in \cite{Talbert:2014bda,Varzielas:2016zuo,Bernigaud:2019bfy}.

However, it does not matter how one finds the parent symmetry with associated RFS for our present purposes.  Any procedure is appropriate, as long as all relevant information about the RFS can be extracted, which we now discuss in detail.

\subsection{Guided Reconstruction for Effective Lagrangians} 
\label{sec:MODEL}
 As demonstrated above, the implied shape of the generators $\hat{T}_a$ given an RFS-invariant Lagrangian is the information required for connecting the IR $\mathcal{G_{\text{a}}}$ to the UV $\mathcal{G_{F,L,Q}}$. But is there a systematic way of using the recovered non-Abelian parent group to build an effective Lagrangian $\mathcal{L}_Y$ (a model, that is) that exhibits, upon family- and electroweak- symmetry breaking, the simplified construction of \eqref{eq:GF}? A straightforward approach\footnote{We do not claim that this is the unique prescription, but rather a simple and economical one.} is based on the implied RFS invariances of $(a)$ the IR mass matrix and $(b)$ new scalar favons whose VEV implement the breaking patterns of \eqref{eq:GF}.

Concerning $(a)$, we note that a generic rotation $V^\dagger_a$ on the mass-eigenstate terms in \eqref{eq:SMfermionmass} yields new mass matrices of the form
\begin{equation}
\label{eq:massgenbasis}
\hat{m}_a \equiv V_{aR} \,m_a\, V^\dagger_{aL} \,\,\,\,\, \Longleftrightarrow \,\,\,\,\, \hat{T}^\dagger_a \, \hat{m}^\dagger_a \hat{m}_a \,\hat{T}_a = \hat{m}^\dagger_a \hat{m}_a = V_{aL} \, m_a^\dagger m_a \, V^\dagger_{aL}\,, 
\end{equation}
where we observed that, by construction, the Hermitian combination of this term is invariant under $\hat{T}_a$ given in \eqref{eq:correspondence}.
Hence, as $m_a$ can always be written as a diagonal matrix of mass eigenvalues, the RFS-invariant quantity $\hat{m}^\dagger_a \hat{m}_a$ can always be written out in model space, once the rotations $V_a$ are specified.  In the class of simplified models we have reviewed above, $V_a$ can always be extracted from the (known) IR phenomenology that is predicted.  In this way \eqref{eq:massgenbasis} provides the rubric for completing the simplified model, as a $\mathcal{L}_Y$ that reproduces it will, by construction, embed the desired RFS. 

Then flavons (point $(b)$) provide a candidate mechanism for breaking $\mathcal{G_{F,L,Q}}$ down to the desired RFS-invariant mass shapes. The $\hat{T}_a$ invariant mass matrices are then obtained, after the flavon expands around its VEV, from Lagrangian terms of the form
\begin{equation}
\label{eq:fullayuka}
\mathcal{L}_Y \supset \frac{\hat{y}_a}\Lambda \, \left[\bar{A}_R\, \phi_a H A_L\right]_{\bf{1}}\,,
\end{equation}
where $A_{L}$ denotes the associated SU(2)$_L$ doublet for the family sector, $A_R$ is the SU(2)$_L$ singlet, $H$ is a Higgs doublet, and $\hat{y}_{a}$ is the effective coupling suppressed by the new physics scale $\Lambda$ integrated out of the effective operator. 
To enforce $\hat{T}_a$ invariant masses, we can use the following condition regarding the VEV direction of the flavon field \cite{Lam:2007qc}
\begin{equation}
\label{eq:vevconstrain}
\hat{T}^\dagger_a \langle \phi_a \rangle = e^\star_a \langle \phi_a \rangle \,\,\, \Longleftrightarrow \,\,\, \hat{T}_a \langle \phi_a \rangle = e_a \langle \phi_a \rangle \,\,\, \,\,\,\,\,\,\,\,\,\text{with} \,\,\,\,\,\,\,\,\, e^\star_a e_a = 1\,,
\end{equation}
where $e_a$ is a (scalar) eigenvalue, and the $\Leftrightarrow$ is due to the fact that $T_a$ is assumed without loss of generality (since we work with finite groups) to be unitary.  From the group theory perspective, it is of course obvious that $\langle \phi_a \rangle$ (or indeed $\langle \phi_a \rangle \cdot \langle \phi_a \rangle$) should preserve $\mathcal{G}_a \cong \hat{T}_a$ when \eqref{eq:vevconstrain} holds. However, in Appendix \ref{sec:APPB} we provide our own derivation of the known condition in \eqref{eq:vevconstrain}, starting from the $\mathcal{G_F}$-invariant Lagrangian term in \eqref{eq:fullayuka}.  This then provides more clarity as to why \eqref{eq:vevconstrain} provides a sufficient condition on operators of the form \eqref{eq:fullayuka} which yields mass matrices of the form in \eqref{eq:massgenbasis}, and thereby realizes the simplified construction in \eqref{eq:GF}.
\subsubsection*{Summary of Effective Lagrangian Reconstruction}
In summary, once a flavour group $\mathcal{G_{F,L,Q}}$ is determined with RFS generated by $\lbrace \hat{T}_a\rbrace$, which are assumed to know about the physical mixings the RFS mechanism (and therefore the reconstructed model) controls, the following procedure can be followed to yield $\mathcal{L}_Y$:
\begin{enumerate}
\item 
\label{step:gens}
Write out the explicit representations of $\lbrace \hat{T}_a\rbrace$ in a basis amenable to manipulating the group product rules of $\mathcal{G_{F,L,Q}}$.  This is typically determined by identifying $\lbrace \hat{T}_a\rbrace$ with specific group elements in a given irreducible representation.  If this basis differs from the one in which $\mathcal{G_{F,L,Q}}$ was originally recovered, take account of the additional transformations in the IR Lagrangian.  This determines the \emph{model basis}.
\item 
\label{step:vevs}
For each family sector $a$ with an active RFS, write down $N_a  \ge 1$ new flavon(s) $\phi_a$ whose VEV respect \eqref{eq:vevconstrain} in the model basis, hence deriving the  model-space orientation of $\langle \phi_a \rangle$. The flavon(s) $\phi_a$ are taken to be charged under the irreducible representation that $\hat{T}_a$ was identified with in Step \ref{step:gens}, and the number of flavon(s) $N_a$ is determined by Steps \ref{step:masses}-\ref{step:operators}, i.e. one needs as many flavons as can successfully yield the desired mass matrix in family sector $a$. 
\item
\label{step:masses}
Derive the expected form of the model-basis mass/coupling matrix in each family sector, which is given by \eqref{eq:massgenbasis} for a generic set of model-basis transformation(s) $V_a$ away from the mass-eigenstate basis, and form the Hermitian combination which preserves the information from the physical mixings in the theory.  
\item 
\label{step:operators}
For each family sector $a$ with an active RFS, create an effective Yukawa-like operator with $\phi_a$ and build invariants of the form in \eqref{eq:fullayuka}, or a similar invariant of the form $\propto \left[L_L L_L \phi_\nu \phi_\nu \right]_{\bf{1}}$ for Majorana neutrinos.
Multiple such invariants may be required in a given family sector, depending on the kinds of irreducible representations implied in Steps \ref{step:gens}-\ref{step:masses}. The goal is to recover the RFS-invariant mass/coupling shapes from Step \ref{step:masses}, with a one-to-one mapping between physical and model parameters, and of course the shapes of $\lbrace \hat{T}_a \rbrace$ and $\hat{m}_a^\dagger \hat{m}_a$ already hint at appropriate generations charges under $\mathcal{G_{F,L,Q}}$.
\item 
\label{step:yukawa}
Construct the Hermitian Yukawa coupling $\hat{Y}_a^\dagger \hat{Y}_a$ from \eqref{eq:fullayuka},\footnote{It is often preferred to instead construct terms in the LR basis, with operators $\propto \bar{A}_L \phi_a H A_R$.  In this case, simply identify the predicted Yukawa coupling from this term as $\hat{Y}_a^\dagger$, and then proceed to build $\hat{Y}_a^\dagger \hat{Y}_a$.  We will do this in some of the models below.  Also, it is obvious that Majorana neutrinos do no require the construction of the Hermitian object $\hat{m}_\nu^\dagger \hat{m}_\nu$.} such that a comparison with the quantities in Step \ref{step:masses} can be made.  Map the model parameters (e.g. $\lbrace \hat{y}_{ai}, v, ... \rbrace$) to physical parameters (e.g. $\lbrace m_{ai}, \theta_{ij}^a, ... \rbrace$).  If this mapping is not one-to-one, the model may appear to require some fine-tuning of parameters, although we will show that this could be a misleading conclusion if the expected RFS-invariant mass matrices have not been generalized with the free parameters permitted through the relationships in \eqref{eq:degenEV}.   Also be sure to check that the implied mass eigenvalues are physical.  If not, additional operators may need to be added.  
\end{enumerate}
If Steps \ref{step:gens}-\ref{step:yukawa} are successful, the resulting model will exhibit the RFS symmetry-breaking patterns and desired phenomenology embedded in the original simplified models, thereby providing an Effective Field Theory (EFT)  completion.

\section{Application to Models of SM Flavour}
\label{sec:SM}
In this section we apply the strategy outlined in Section \ref{sec:MODEL} to flavour models reproducing SM mixing matrices, i.e. the PMNS or CKM matrices defined in \eqref{eq:CKMPMNS}.  

\subsection{$A_4$ Altarelli-Feruglio Model for $U_{PMNS}$}
\label{sec:A4AF}
As a first application of the algorithm described in Section \ref{sec:THEORY}, we now show how the famous Altarelli-Feruglio model of leptonic flavour \cite{Altarelli:2005yx,Altarelli:2010gt} can be reconstructed with only minimal knowledge of its low-energy predictions. In particular, its IR phenomenology is characterized by the breaking of the tetrahedral $A_4$ group to $\mathbb{Z}_3$ and $\mathbb{Z}_2$ RFS in the charged lepton and neutrino sectors, 
\begin{equation}
\label{eq:A4AF}
\mathcal{G_{L}} \cong A_4   \rightarrow	 \begin{cases} 
							
				            		 \mathcal{G_{\nu}} \cong \mathbb{Z}_2
					     		 \\
					     		 \mathcal{G_{\text{l}}} \, \cong \mathbb{Z}_3 \,\,,
							
						\end{cases} \\
\end{equation}
which are respectively generated by 
\begin{equation}
\label{eq:GensAF}
T_{\nu} = \text{diag} \left(-1, 1, -1 \right), \,\,\,\,\,\,\,\,\,
T_l = \text{diag} \left(1, \omega^2, \omega \right), \,\,\,\,\,\,\,\,\,\, \omega \equiv e^{2 \pi i/3} \,.
\end{equation}
The model assumes Majorana neutrino masses, and its LO mixing prediction is the tribimaximal (TBM) matrix defined as follows \footnote{In this particular example, our convention for the PMNS mixing is different than that adopted later in the paper, cf. \eqref{eq:mt2}, in order to better reproduce those of \cite{Altarelli:2005yx}.} 
\begin{equation}
\label{eq:PMNSAF}
U_{PMNS}  \equiv \left(
\begin{array}{ccc}
\sqrt{2/3} & 1/\sqrt{3} & 0 \\
-1/\sqrt{6} & 1/\sqrt{3} & -1/\sqrt{2} \\
-1/\sqrt{6} & 1/\sqrt{3} & 1/\sqrt{2}
\end{array}
\right)\,,
\end{equation}
which is realized in a special flavour basis, where $U_l = \mathbb{1}$ and $U_\nu \equiv U_{PMNS}$.\footnote{Note that one is always free to move to this basis, and so knowledge of this fact about the Altarelli-Feruglio model does not bias our approach in what follows.}  Given knowledge of \eqref{eq:GensAF}-\eqref{eq:PMNSAF}, we now have all of the information necessary to apply our reconstruction algorithm. 

We first use these equations to infer that, in the model basis we will construct our effective Lagrangian, the relevant generators are given by 
\begin{equation}
\label{eq:GensAFFB}
T_{\nu U} = \frac{1}{3}\left(
\begin{array}{ccc}
-1 & 2 & 2 \\
2 & -1 & 2 \\
2 & 2 & -1
\end{array}
\right), \,\,\,\,\,
T_{lU} = \left(
\begin{array}{ccc}
1 & 0 & 0 \\
0 & \omega^2 & 0 \\
0 & 0 & \omega
\end{array}
\right)\,,
\end{equation}
which we immediately identify as triplet ${\bf{3}}$ representations from the $A_4$ review in \cite{Altarelli:2010gt}, and which we can use to solve for flavon VEV in each family sector,
\begin{equation}
\label{eq:flavonVEVAF}
T_{\nu U,lU} \, \langle \phi_{\nu,l} \rangle = \langle \phi_{\nu,l} \rangle  \,\,\, \Longrightarrow \,\,\, \langle \phi_\nu \rangle = \left(
\begin{array}{c}
v_\nu \\
v_\nu \\
v_\nu
\end{array}
\right),
\,\,\,\,\, \langle \phi_l \rangle = \left(
\begin{array}{c}
v_l \\
0 \\
0
\end{array}
\right)\,,
\end{equation} 
and to conclude that the model-basis mass matrices invariant under them are characterized by
\begin{align}
\label{eq:mnuAFinfer}
m_{\nu U} &\overset{!}{=} \frac{1}{3} \left(
\begin{array}{ccc}
 \left( 2m_{\nu_1} + m_{\nu_2} \right) & \left(m_{\nu_2} - m_{\nu_1}\right) & \left( m_{\nu_2} - m_{\nu_1}\right) \\
\left(m_{\nu_2} - m_{\nu_1}\right) & \frac{1}{2} \left( 3 m_{\nu_3} + 2 m_{\nu_2} + m_{\nu_1}\right) & \frac{1}{2} \left(m_{\nu_1} +2 m_{\nu_2} - 3 m_{\nu_3}  \right) \\
 \left( m_{\nu_2} - m_{\nu_1}\right) &\frac{1}{2} \left(m_{\nu_1} +2 m_{\nu_2} - 3 m_{\nu_3}  \right) & \frac{1}{2} \left( 3 m_{\nu_3} + 2 m_{\nu_2} + m_{\nu_1}\right)
\end{array} 
\right), \\
\label{eq:mLAFinfer}
m_{lU}^\dagger m_{lU} &\overset{!}{=} \left(
\begin{array}{ccc}
m_{e}^2 & 0 & 0 \\
0 & m_\mu^2 & 0 \\
0 & 0 & m_\tau^2
\end{array}
\right),
\end{align}
which we can further use to infer charge assignments under $A_4$ for SM fields and the new flavons in \eqref{eq:flavonVEVAF}.  We recall that $A_4$ is $\mathcal{O}(12)$ and has four irreducible representations:  a triplet $\bf{3}$ and three singlets $\bf{1}$, $\bf{1^\prime}$ and $\bf{1^{\prime\prime}}$, with $\bf{1}$ denoting the trivial representation.  While $A_4$ is an exceptionally well studied finite group, we repeat the relevant product rules in this basis for completeness in Appendix \ref{app:A4}, where it is clear that in order to build up non-trivial Yukawa matrices, the SM SU(2) doublet $L_L$ and corresponding flavons (from \eqref{eq:flavonVEVAF}) will need to be assigned to the triplet representation:
\begin{equation}
\label{eq:tripletfieldsAF}
L_L \sim \bf{3}, \,\,\,\,\,\,\,\,\,\, \phi_\nu \sim \bf{3}, \,\,\,\,\,\,\,\,\,\, \phi_l \sim \bf{3} \,.
\end{equation}
Given this, we then consider the charged lepton mass term and observe from \eqref{eq:GensAFFB} and \eqref{eq:mLAFinfer} that SM generations do not `talk' to one another through the $A_4$ symmetry, and so we assign a different singlet to each RH SM field:
\begin{equation}
\label{eq:singletfieldsAF}
e_{R}^c \sim {\bf{1}} \,, \,\,\,\,\,\,\,\,\,\,\mu_{R}^c \sim {\bf{1^{\prime \prime}}} \,,\,\,\,\,\,\,\,\,\,\,\tau_{R}^c \sim {\bf{1^{\prime}}}\,,
\end{equation}
where $l_{R}^c$ are transforming as left-handed fields. Because these fields transforms as $A_4$ singlets we need combinations of $\left[\phi_l L_L \right]$ as in \eqref{eq:fullayuka} to themselves transform as one-dimensional representations of $A_4$.  Noting this, one quickly deduces the LO effective Yukawa Lagrangian for this sector:
\begin{equation}
\label{eq:LlYAF}
\mathcal{L}_{l} \supset a_e \, e_{R}^c \, \left[ \phi_l L_L \right]_{\bf{1}} + b_e \, \mu_{R}^c \left[ \phi_l L_L\right]_{\bf{1^\prime}} + c_e \, \tau_{R}^c \left[ \phi_l L_L \right]_{\bf{1^{\prime\prime}}}\, + \text{h.c} + \text{...},
\end{equation}
where we omit the necessary insertions of the Higgs field that make each term invariant and the $\left[ ... \right]_{\bf{1^\prime}}$ notation indicates that the bracketed fields contract to the indicated singlet under the $A_4$ product rules given in \eqref{eq:singletproductA4}.  Each individual term in \eqref{eq:LlYAF} is then an $A_4$ and SM gauge singlet, once contracted with the corresponding RH isospin singlets.  The additional terms implied in \eqref{eq:LlYAF} correspond to higher-order operators in the Effective Field Theory (EFT) allowed by successive flavon and SM field insertions, given their associated symmetry assignments.  We will discuss these below, along with additional symmetries irrelevant to the RFS formalism.  Regardless, one immediately finds that \eqref{eq:LlYAF} generates the desired mass matrix from \eqref{eq:mLAFinfer}, with the relations between masses and Lagrangian parameters easily found to be
\begin{equation}
m_e = a_e\,v\,\frac{v_l}{\Lambda}, \,\,\,\,\,m_\mu = b_e\,v\,\frac{v_l}{\Lambda},\,\,\,\,\,m_\tau = c_e\,v\,\frac{v_l}{\Lambda},
\end{equation}
with $v$ the Higgs VEV realizing EWSB.

\begin{table}[t]
    \centering
    \renewcommand{\arraystretch}{1.5}
    \begin{tabular}{|c||c|c|c|c|c|c|c|c|c|c|c|}
    \hline
  \  & $L_L$ &  $e_{R}^c$ & $\mu_{R}^c$ & $\tau_{R}^c$ & $\phi_\nu$ & $\phi_l$ & $\xi$ \\
  \hline
  \hline
  $A_4$ & $\mathbf{3}$ & $\mathbf{1}$ & $\mathbf{1^{\prime\prime}}$ & $\mathbf{1^\prime}$ & $\mathbf{3}$ & $\mathbf{3}$ & $\mathbf{1}$ \\
  \hline
\end{tabular}    
\caption{Relevant field and $A_{4}$ symmetry content from \cite{Altarelli:2005yx}.}
    \label{Tab:A4QuantumNumbers}
\end{table}

Moving now to the neutrino masses, the $m_{\nu U}$ implied in \eqref{eq:mnuAFinfer} has non-trivial structure in all matrix sectors, a fact concurrent with $1)$ our observation that $\phi_\nu$ and $L_L$ should be charged as $A_4$ triplets, and $2)$ the fact the Altarelli-Feruglio Model predicts a Majorana neutrino mass matrix, which is itself implied (or at least consistent with) the $\mathbb{Z}_2$ neutrino RFS.  In the low-energy EFT, a Majorana neutrino mass is necessarily $\propto L_L L_L$.  We therefore conclude that an operator of the form\footnote{Note that a see-saw realzation of this IR Majorana term is also possible --- see \cite{Altarelli:2005yx}.}
\begin{equation}
\label{eq:LnuYAFa}
\mathcal{L}_\nu \supset a_\nu \left[\phi_\nu L_L L_L \right]_{\bf{1}} + \text{h.c} + \text{...},
\end{equation}
should be included in the Lagrangian.  This term generates a contribution to $m_{\nu U}$,
\begin{equation}
\label{eq:cont1mNUAF}
m_{\nu U} \simeq \frac{v^2}{\Lambda} \left(
\begin{array}{ccc}
2B/3 & -B/3 & -B/3 \\
-B/3 & 2B/3 & -B/3 \\
-B/3 & -B/3 & 2B/3
\end{array}
\right), \,\,\,\,\,\,\,\,\, B \equiv 2 a_\nu \frac{v_\nu}{\Lambda} \,,
\end{equation}
which, while invariant under \eqref{eq:GensAFFB} (as it must be by construction), fails to realize the required neutrino phenomenology, as it has only two distinct eigenvalues: $0$ and $B$.  In other words, it cannot map to the generic, RFS-invariant form in \eqref{eq:mnuAFinfer} that we have deduced, in the absence of (unphysical) assumptions about the mass eigenvalues embedded in it.  The obvious solution is to introduce a further flavon $\xi$ whose VEV $\langle \xi \rangle = u$ does not break $\mathcal{G}_\nu$ and which can couple to the $L_L L_L$ bilinear.  To that end we introduce $\xi$ as an $A_4$ singlet, which adds an additional contribution to \eqref{eq:LnuYAFa}-\eqref{eq:cont1mNUAF},
\begin{equation}
\label{eq:LnuYAFb}
\mathcal{L}_\nu \supset  a_\nu \left[\phi_\nu L_L L_L \right]_{\bf{1}} \, + \,b_\nu \xi \left[ L_L L_L \right]_{\bf{1}}  \Longrightarrow m_{\nu U} \simeq \frac{v^2}{\Lambda} \left(
\begin{array}{ccc}
A + 2B/3 & -B/3 & -B/3 \\
-B/3 & 2B/3 & A - B/3 \\
-B/3 & A - B/3 & 2B/3
\end{array}
\right), \,\,\,\, A \equiv 2 b_\nu \frac{u}{\Lambda} \,,
\end{equation}
where we again omit the necessary insertions of the Higgs field that make each term invariant.
This matrix is still invariant under \eqref{eq:GensAFFB}, is diagonalized by $U_{TBM}$, and has mass eigenvalues given by  
\begin{equation}
\label{eq:AFev}
m_{\nu_1} = \frac{v^2}{\Lambda} \left(A + B\right), \,\,\,\,\,\,\,\,\,\,m_{\nu_2} = \frac{v^2}{\Lambda} A, \,\,\,\,\,\,\,\,\,\,m_{\nu_3} = \frac{v^2}{\Lambda} \left(B - A\right),
\end{equation}
which is fully consistent with the matrix form in \eqref{eq:mnuAFinfer}.

In conclusion, with the knowledge of the parent flavour symmetry $A_4$, the neutrino and charged lepton RFS in \eqref{eq:GensAF}, and the PMNS mixing prediction given in \eqref{eq:PMNSAF}, we have easily inferred the field and symmetry content in Table \ref{Tab:A4QuantumNumbers} and the following LO effective Yukawa Lagrangian:
\begin{equation}
\label{eq:AFLagfull}
\mathcal{L}_Y \supset a_e \, e_{R}^c \, \left[ \phi_l L_L \right]_{\bf{1}}+ b_e \, \mu_{R}^c \left[ \phi_l L_L\right]_{\bf{1^\prime}} + c_e \, \tau_{R}^c \left[ \phi_l L_L \right]_{\bf{1^{\prime\prime}}} + a_\nu \left[\phi_\nu L_L L_L \right]_{\bf{1}} + b_\nu \xi \left[ L_L L_L \right]_{\bf{1}}+ \text{h.c} + \text{...}.
\end{equation}
This is to be compared to eq(12) in \cite{Altarelli:2005yx}, where it is found to be equivalent to the non-SUSY version of the Altarelli-Feruglio Lagrangian --- we have `reconstructed' this model from the bottom up.

Here we do not concern ourselves with the UV completion of this model (or indeed other models), which can be achieved by adding appropriate messengers fields to make the underlying model renormalizable, for $A_4$ models see e.g. \cite{Varzielas:2010mp,Varzielas:2012ai}. UV completions exist in general for models and are typically more predictive than the corresponding non-renormalizable model if the messenger fields included in the complete model enable a subset of the contractions that are allowed by the symmetries at the non-renormalizable level.

However, as is well known, the complete model of \cite{Altarelli:2005yx} is more involved than just its LO Yukawa terms.  Furthermore, we made choices in the above discussion that, a priori, may seem ad-hoc. We will now discuss some of these subtleties for this particular model, as well as their broader implications for our generic approach, although in forthcoming models we will typically leave these discussions implicit, unless they become particularly relevant for the physics at hand.

\subsubsection*{Mass and Mixing Prediction Ambiguities}
We have observed in the preceding section that knowledge of the IR RFS and mixing prediction is not guaranteed to tell us \emph{everything} required to build the LO terms in the EFT.   For one, as became clear between \eqref{eq:cont1mNUAF}-\eqref{eq:LnuYAFb}, the RFS has no control over the quantization of the mass eigenvalues, but only the mixing associated to them.\footnote{This is a well known fact about RFS that are only active in SM mass terms.}  As we saw, \eqref{eq:cont1mNUAF} exhibits the required $\mathbb{Z}_2$ invariance, but does not map to the (more) generic RFS-invariant form in \eqref{eq:mnuAFinfer} without imposing,
\begin{equation}
m_{\nu_1} \rightarrow m_{\nu_3}, \,\,\,\,\,\,\,\,\,\,m_{\nu_2} \rightarrow 0,
\end{equation}
which is unphysical (there are two non-zero mass splittings measured for low-energy neutrinos).  This motivated the introduction of the singlet $\xi$ whose VEV also breaks $\mathcal{G_L}$, but does \emph{not} break $\mathcal{G_\nu}$, and so does not upset the TBM prediction for the PMNS.  In general, this is a good strategy when reconstructing a given Majorana-neutrino-sector Lagrangian for which one does not have a literature reference, as we did here ---- operators $\propto \left[L_L L_L\right]_{\bf{1}}$ will always preserve a given RFS if augmented only by a scalar singlet.  

And secondly, in the absence of a reference Lagrangian, one can also reconstruct a mass term associated to a different mixing prediction, when degenerate eigenvalues exist in RFS generators.  This was highlighted explicitly for the mixing in \eqref{eq:degenEV}, but of course this also has implications for the associated RFS-invariant mass matrix.  In the Altarelli-Feruglio case, we observe that $T_l$ has three distinct eigenvalues, and so $U_l$ is uniquely predicted as the identity matrix.  However, as only one $\mathbb{Z}_2$ is explicitly preserved in the neutrino sector, the truly generic RFS-invariant mass matrix is given by
\begin{align}
\label{eq:genM13AF}
m_{\nu U}^{11} &= \frac{1}{3} \left(m_{\nu_2} + 2m_{\nu_1}\cos^2 \theta_{13} + 2 m_{\nu_3} e^{2 i \delta_{13}} \sin^2 \theta_{13}  \right),  \\
\nonumber
m_{\nu U}^{12} 
&= \frac{1}{3} \left(m_{\nu_2} - m_{\nu_1} \cos^2\theta_{13} + \sqrt{3} e^{-i\delta_{13}} \left(m_{\nu_1} - e^{2i\delta_{13}} m_{\nu_3} \right) \cos \theta_{13} \sin \theta_{13} - m_{\nu_3} \sin^2\theta_{13}e^{2 i\delta_{13}}   \right),  \\
\nonumber
m_{\nu U}^{13} 
&=\frac{1}{3} \left(m_{\nu_2} - m_{\nu_1} \cos^2\theta_{13} - \sqrt{3} e^{-i\delta_{13}} \left(m_{\nu_1} - e^{2i\delta_{13}} m_{\nu_3} \right) \cos \theta_{13} \sin \theta_{13} - m_{\nu_3} \sin^2\theta_{13}e^{2 i\delta_{13}}   \right),    \\
\nonumber
m_{\nu U}^{22} &= \frac{1}{18} \left(6 m_{\nu_2} + m_{\nu_1} \left(\sqrt{3} \cos \theta_{13} - 3 e^{-i \delta_{13}} \sin \theta_{13} \right)^2 + m_{\nu_3} \left(3 \cos \theta_{13} + \sqrt{3} e^{i \delta_{13}} \sin \theta_{13} \right)^2 \right),  \\
\nonumber
m_{\nu U}^{23} 
&= \frac{1}{6} \left(2 m_{\nu_2} + \left(m_{\nu_1} - 3 m_{\nu_3} \right) \cos^2 \theta_{13} + e^{-2 i \delta_{13}} \left(-3 m_{\nu_1} + e^{4i \delta_{13}} m_{\nu_3} \right)\sin^2 \theta_{13}\right),  \\
\nonumber
m_{\nu U}^{33} &= \frac{1}{18} \left(6m_{\nu_2} + e^{-2 i \delta_{13}} m_{\nu_1} \left(\sqrt{3} e^{i \delta_{13}} \cos \theta_{13} + 3 \sin \theta_{13} \right)^2 + m_{\nu_3} \left(-3 \cos \theta_{13} + \sqrt{3} e^{i \delta_{13}} \sin \theta_{13} \right)^2 \right), 
\end{align}
with $\theta_{13}$ and $\delta_{13}$ defined as in \eqref{eq:degenEV}.  This complex symmetric matrix is diagonalized by $U_{TBM-13}^T \cdot m_{\nu U} \cdot U_{TBM-13} = m_{\nu}$, with
\begin{equation}
\label{eq:TBM13}
U_{TBM-13} = \left(
\begin{array}{ccc}
\sqrt{2/3} \cos \theta_{13} & 1/\sqrt{3} & \sqrt{2/3}\, e^{-i \delta_{13}} \sin \theta_{13} \\
-\frac{\cos \theta_{13}}{\sqrt{6}} + \frac{e^{i \delta_{13}}\,\sin \theta_{13}}{\sqrt{2}} & 1/\sqrt{3} & -\frac{\cos \theta_{13}}{\sqrt{2}} - \frac{e^{-i \delta_{13}}\,\sin \theta_{13}}{\sqrt{6}}  \\
-\frac{\cos \theta_{13}}{\sqrt{2}} - \frac{e^{-i \delta_{13}}\,\sin \theta_{13}}{\sqrt{6}} & 1/\sqrt{3} &  \frac{\cos \theta_{13}}{\sqrt{2}} - \frac{e^{-i \delta_{13}}\,\sin \theta_{13}}{\sqrt{6}} 
\end{array}
\right)\,,
\end{equation}
from which $\theta_{13}$ and $\delta_{13}$ can be fit to experimental data, yielding a phenomenologically successful PMNS matrix.
The point here is that knowledge of the RFS alone does not nail down the mixing or mass matrix prediction that a top-down EFT can yield, when the generators of said RFS do not distinguish all three generations.  That the Altarelli-Feruglio model predicts \eqref{eq:PMNSAF} and not \eqref{eq:TBM13} at leading order is due to the accidental invariance of the $m_{\nu U}$  in \eqref{eq:LnuYAFb} under the $\mu-\tau$ operator\footnote{Note that, using the basis choice for the Klein RFS generating set in \eqref{eq:SMRFS}, one would instead derive the matrix $-T_{\mu-\tau}$ from the bottom-up. which still leaves $m_{\nu U}$ invariant but corresponds to a different $\hat{e}_\nu$.} 
\begin{equation}
\label{eq:mutau}
T_{\mu-\tau} = \left(
\begin{array}{ccc}
1 & 0 & 0 \\
0 & 0 & 1 \\
0 & 1 & 0
\end{array}
\right)\,,
\end{equation}
which \eqref{eq:genM13AF} does not respect. This invariance is not associated to the RFS, as $\mathbb{Z}_2 \times \mathbb{Z}^{\mu \tau}_2$ is not a subgroup of $A_4$.
Rather, the accidental invariance of \eqref{eq:LnuYAFb} under \eqref{eq:mutau} is due to the absence of additional operators in \eqref{eq:AFLagfull}, which is a consequence of additional symmetries unrelated to the RFS, which we will now discuss. 
\subsubsection*{Further Symmetries and Fields}
The astute reader will notice that \eqref{eq:AFLagfull} does not contain the most generic set of operators invariant under the $A_4$ flavour symmetry and SM gauge symmetries.  For example, a term of the form (briefly restoring Higgs fields to maintain clarity)
\begin{equation}
\mathcal{O} \sim L_L L_L H H 
\end{equation}
is also allowed, as are the four operators corresponding to \eqref{eq:AFLagfull} but with $\phi_\nu \leftrightarrow \phi_l$.  Indeed, these additional contributions to the LO Lagrangian are forbidden by a $\mathbb{Z}_3$ \emph{shaping} symmetry, 
\begin{equation}
\lbrace e, \mu, \tau \rbrace_R \sim \omega^2, \,\,\,\,\, \lbrace H, \phi_{l}\rbrace \sim 1, \,\,\,\,\, \lbrace L_L, \xi, \phi_{\nu}\rbrace \sim \omega \,,
\end{equation}
which limits contact interactions between certain fields.  As the RFS have nothing to do with these shaping symmetries, in what follows we will simply assume that either they are not needed or, more commonly, that they can always be found such that only desired operators in the $1/\Lambda$ EFT expansion are recovered.

We also ignored all of the dynamics required to obtain the VEV derived in \eqref{eq:flavonVEVAF}.  As mentioned in the introduction, flavon VEV can be realized via the minimization of an appropriate scalar potential.  SUSY is assumed for the Altarelli-Feruglio model, such that \eqref{eq:AFLagfull} is understood as one part of the overall superpotential, whilst yet another flavon $\tilde{\xi}$  that breaks $A_4$ is introduced alongside additional `driving' superfields $\phi^{l}_0$, $\phi^\nu_0$ and $\xi_0$.  All fields are then further charged under a traditional $R$ symmetry U(1)$_R$ that distinguishes matter, symmetry-breaking, and driving/alignment fields.  We again ignore all such discussion in upcoming models, as we simply assume that the required VEV alignment can be achieved. 

Finally, we mentioned that the RFS do not constrain mass eigenvalues.  That means that mass hierarchies must be understood with some other mechanism. In the case of \cite{Altarelli:2005yx}, this is achieved with an additional Froggatt-Nielsen \cite{Froggatt:1978nt} U(1)$_{FN}$, under which the $\tau_R$, $\mu_R$, and $e_R$ generations are assigned $0$, $q$ and $2q$, and additional flavons $\theta$ are introduced whose VEV create hierarchical mass suppressions $\sim \lambda \equiv \langle \theta \rangle / \Lambda$:  $c_e \approx \mathcal{O}(1)$, $b_e \approx \mathcal{O}(\lambda^q)$, and $a_e \approx \mathcal{O}(\lambda^{2q})$.  Again, such symmetries can always be imposed in addition to the core flavour symmetries yielding the RFS of interest.  We therefore do not mention them further in what follows.

\subsubsection*{Higher-order Operators}
We have only reconstructed the LO Yukawa Lagrangian in the $1/\Lambda$ EFT expansion.  Higher-order terms associated to more SM or flavon field insertions (but which are still invariant under \emph{all} assumed symmetries) can of course be found, and these will generate small corrections to the phenomenological conclusions of the LO Lagrangian.  In the Altarelli-Feruglio model, the leading such terms are given by 
\begin{equation}
\label{eq:HOAFl}
\mathcal{O} \sim \left[e_R L_L \phi_l \phi_l \right]_{\bf{1}}, \,\,\,\,\,\mathcal{O} \sim \left[\mu_R L_L \phi_l \phi_l \right]_{\bf{1}}, \,\,\,\,\, \mathcal{O} \sim \left[\tau_R L_L \phi_l \phi_l \right]_{\bf{1}}, 
\end{equation}
for the charged leptons, and 
\begin{equation}
\label{eq:HOAFl}
\mathcal{O} \sim \left[\phi_l \phi_\nu\right]_{\bf{1^\prime}} \left[L_L L_L\right]_{\bf{1^{\prime\prime}}}, \,\,\,\,\,\mathcal{O} \sim \left[\phi_l \phi_\nu\right]_{\bf{1^{\prime\prime}}} \left[L_L L_L\right]_{\bf{1^\prime}}, \,\,\,\,\, \mathcal{O} \sim \xi\left[\phi_l L_L L_L \right]_{\bf{1}}, 
\end{equation}
for the neutrinos.  These operators will add small corrections to the predictions for the associated mixing matrices, which can bring them closer to experiment.  However, in general, they may softly break the RFS preserved at LO,\footnote{Combinations of flavons generally give rise to different directions in flavour space, so-called effective alignments \cite{deMedeirosVarzielas:2017hen, deMedeirosVarzielas:2019hur}.} and so studying them in generality is again beyond our scope in what follows.

\subsection{$\left(\mathbb{Z}_{14} \times \mathbb{Z}_2 \right) \rtimes \mathbb{Z}_{2}$ Model for $U_{CKM}$}
\label{sec:CompSM}

We now consider a simple model based on the finite group $\left(\mathbb{Z}_{14} \times \mathbb{Z}_2 \right) \rtimes \mathbb{Z}_{2}$ that makes predictions for CKM quark mixing.  The RFS symmetry-breaking pattern to the down and up quark sectors is illustrated by
\begin{equation}
\label{eq:Compdiag}
\mathcal{G_{Q}} \cong \left(\mathbb{Z}_{14} \times \mathbb{Z}_2 \right) \rtimes \mathbb{Z}_{2}  \rightarrow	 \begin{cases} 
							
				            		 \mathcal{G_{\text{u}}} \cong \mathbb{Z}_2
					     		 \\
					     		 \mathcal{G_{\text{d}}} \, \cong \mathbb{Z}_2 \,\,,
							
						\end{cases} \\
\end{equation}
with $\mathcal{G_{\text{u,d}}}$ generated by
\begin{equation}
T_u = \text{diag}\left(1, -1, -1\right), \,\,\,\,\,\,\,\,\,\,T_d = \text{diag}\left(1, -1, -1\right)\,.
\end{equation}
The model predicts the LO CKM mixing prediction to be of the Cabibbo form,
\begin{equation}
\label{eq:CKMComp}
U_{CKM} = 
\left(
\begin{array}{ccc}
\cos \frac{\pi}{14} & \sin \frac{\pi}{14} & 0  \\
- \sin \frac{\pi}{14} & \cos \frac{\pi}{14} & 0    \\
0   & 0 & 1
\end{array}
\right)\, , 
\end{equation}
which, while insufficient to fully reproduce the known three-dimensional structure of the CKM, does capture the dominant mixing in the $(1,2)$ sector, i.e. the Cabibbo angle $\theta_ C$.  Further corrections are highly suppressed, $\propto \mathcal{O}\left( \theta_{C}^{2}, \theta_{C}^{3} \right)$, and will be briefly mentioned below.

As in Section \ref{sec:A4AF}, we can immediately construct the flavour-basis generators, under the (common) assumption that the down quarks are already diagonal, such that the entirety of the CKM mixing is encoded in the up sector.  We immediately find
\begin{equation}
\label{eq:Compflavbasgen}
T_{uU} = \left(
\begin{array}{ccc}
\cos \frac{\pi}{7} & \sin \frac{\pi}{7} & 0\\
\sin \frac{\pi}{7} & -\cos \frac{\pi}{7}& 0\\
0 & 0 & -1
\end{array}
 \right)\,, \,\,\,\,\,\,\,\,\,\, T_{dU} = \left(
 \begin{array}{ccc}
1 & 0 & 0\\
0 & -1 & 0\\
0 & 0 & -1
 \end{array}
 \right)\,.
\end{equation}
In principle, we can use \eqref{eq:Compflavbasgen} to proceed with the algorithm as described in Section \ref{sec:THEORY}. However, in what follows we find it convenient to work in a different basis where the non-trivial entries of $T_{uU}$ are in the $(2,3)$ sector.\footnote{We also attempted a construction with this symmetry that treated leptons as well as quarks, which we reference at the end of Section \ref{sec:D15}, and where the current basis was required.}  To that end, we consider the following unitary transformation on the weak-eigenstate generators: 
\begin{equation} 
\label{eq:Comppmatrix}
P =
\frac{1}{2} \begin{pmatrix}
1 &1 + \frac{i}{\sqrt{2}} & 1 - \frac{i}{\sqrt{2}} \\
-1 & 1 - \frac{i}{\sqrt{2}} & 1 + \frac{i}{\sqrt{2}} \\
\sqrt{2} & -i & i
\end{pmatrix} .
\end{equation}
Applying $P$ with $T_{u,d}^{\prime \prime} = P^\dagger T_{uU,dU} P$ we get the following expressions for the RFS-generators in the model basis:\footnote{In what follows, we will always use double-primed ($\prime\prime$) notation when constructing objects in the basis we intend to build the model, unless that basis has already been given a specific label (as with the flavour basis in Section \ref{sec:A4AF}).}
\begin{align}
\begin{split}
T_{u}^{\prime \prime} 
= \begin{pmatrix}
-1 & 0 & 0 \\
0 & 0 & e^{i \pi/7} \\
0 &e^{-i \pi/7} &0
\end{pmatrix}, \,\,\,\,\,
T_{d}^{\prime \prime} 
= \begin{pmatrix}
-1 & 0 & 0\\
0 & 0 & 1\\
0 & 1 & 0
\end{pmatrix}\,.
\end{split}
\label{Left_RFS_in_Model_Comp}
\end{align}
The previous expressions \eqref{Left_RFS_in_Model_Comp} indicate the use of a singlet and doublet representation. Even if the second and third generation are part of the doublet, we remind the reader that these generations are not the actual flavor states. In principle it is possible to build the model in the flavour basis, but the charge assignments would be rather inconvenient. Finally, let us emphasize that the actual flavor state charges do not depend of the group basis choice.

We now want to identify these matrices with $\left(\mathbb{Z}_{14} \times \mathbb{Z}_2 \right) \rtimes \mathbb{Z}_{2}$ generating elements in certain irreducible representations. However, we have not found sources available that catalogue the properties of this group.  For this reason we derived the relevant product rules and group information ourselves, and have provided them in Appendix \ref{app:complicated}. There we see that $T_{u,d}^{\prime \prime}$ can be easily expressed in terms of the group generators ${\bf{a}}$, ${\bf{b}}$ and ${\bf{c}}$: 
\begin{equation}
\begin{split}
&T_{d}^{\prime \prime} = {\bf{bab}}, \ \ \ T_{u}^{\prime \prime} = {\bf{ac}} \ \ \ \ \ \  \text{for} \ \ \ \ \ \mathbf{1_{--}} +\mathbf{2_{3+-}}\,. 
\end{split}
\label{Preserved_elements_Comp}
\end{equation}
Critically, we observe that \eqref{Left_RFS_in_Model_Comp}-\eqref{Preserved_elements_Comp} indicate that $\mathbf{2_{3+-}}$ is the appropriate $\left(\mathbb{Z}_{14} \times \mathbb{Z}_2 \right) \rtimes \mathbb{Z}_{2}$ charge for the two flavons $\phi_{u,d}$ that we introduce according to the algorithm in Section \ref{sec:THEORY}, and we can use \eqref{Left_RFS_in_Model_Comp} to work out the expressions for these doublet VEV, finding
\begin{equation}
\ \ \ \langle \phi_u \rangle =v_u \left(
\begin{array}{c}
1 \\
-e^{-i \pi/7}
\end{array}
\right), \ \ \ \langle \phi_{d} \rangle = v_d \left(
\begin{array}{c}
1 \\
-1
\end{array}
\right) \, .
\label{compvevs}
\end{equation}
Finally, one derives that in the absence of mixing ambiguities, the model-basis mass matrices are given by 
\begin{align}
\nonumber
m_{u}^{''\dagger}m_{u}^{''}&\overset{!}{=}
\left( 
\begin{array}{ccc}
m_{u_3}^2 & 0 & 0 \\
0 & \frac{1}{2} \left(m_{u_1}^2 + m_{u_2}^2 \right) &\frac{1}{2}e^{i \pi/7} \left(m_{u_1}^2 - m_{u_2}^2\right)\\
0 & \frac{1}{2}e^{-i\pi/7} \left(m_{u_1}^2 - m_{u_2}^2\right) & \frac{1}{2} \left(m_{u_1}^2 + m_{u_2}^2 \right)
\end{array}
\right), \\
\label{eq:CompgenMass}
m_{d}^{''\dagger}m_{d}^{''}&\overset{!}{=}
\left( 
\begin{array}{ccc}
m_{d_3}^2 & 0 & 0 \\
0 & \frac{1}{2} \left(m_{d_1}^2 + m_{d_2}^2\right) & \frac{1}{2} \left(m_{d_1}^2 - m_{d_2}^2\right) \\
0 & \frac{1}{2} \left(m_{d_1}^2 - m_{d_2}^2\right) & \frac{1}{2} \left(m_{d_1}^2 + m_{d_2}^2\right)
\end{array}
\right),
\end{align}
where $m_{A_i}$ are the associated mass eigenvalues, and where we have used dagger combinations for the charged fermions to remove the dependence on RH transformations.  Of course, in deriving \eqref{eq:CompgenMass}, we have been careful to keep track of the additional basis change implied by operating with $P$ in \eqref{eq:Comppmatrix}.

The results in \eqref{Left_RFS_in_Model_Comp}-\eqref{eq:CompgenMass} strongly indicate that the second and third generations of LH quarks should transform as a $\left(\mathbb{Z}_{14} \times \mathbb{Z}_2 \right) \rtimes \mathbb{Z}_{2}$ doublet, while the first generation transforms as a non-trivial singlet.  Similarly, the second and third generations of RH up and down quarks should transform as a non-trivial singlet, while the first generation of both types of quark transforms trivially. Furthermore, \eqref{Preserved_elements_Comp} already indicated that the flavons $\phi_{d,u}$ associated to these sectors should transform as a $\mathbf{2_{3+-}}$, a fact that helped us derive \eqref{compvevs}.  This information is summarized in Table \ref{Tab:CompQuantumNumbers}. 
\begin{table}[t]
    \centering
    \renewcommand{\arraystretch}{1.5}
    \begin{tabular}{|c||c|c|c|c|c|c|c|c|c|c|c|}
    \hline
  \  & $Q_L^{''1}$ & $Q_L^{''23}$ & $u_{R}^{''1}$ & $u_{R}^{''2}$ & $u_{R}^{''3}$ & $d_{R}^{''1}$ & $d_{R}^{''2}$ & $d_{R}^{''3}$  & $\phi_u$ & $\phi_d$ \\
  \hline
  \hline
  $\left(\mathbb{Z}_{14} \times \mathbb{Z}_2 \right) \rtimes \mathbb{Z}_{2}$ & $\mathbf{1_{--}}$ & $\mathbf{2_{3+-}}$ & $\mathbf{1_{++}}$ & $\mathbf{1_{-+}}$ & $\mathbf{1_{--}}$ & $\mathbf{1_{++}}$ & $\mathbf{1_{-+}}$ & $\mathbf{1_{--}}$ & $\mathbf{2_{3+-}}$ & $\mathbf{2_{3+-}}$ \\
  \hline
\end{tabular}    
\caption{Representations of the quarks under $\left(\mathbb{Z}_{14} \times \mathbb{Z}_2 \right) \rtimes \mathbb{Z}_{2}$.}
    \label{Tab:CompQuantumNumbers}
\end{table}

Assuming a shaping symmetry to prevent $\phi_i$ from coupling to undesirable sectors, one can straightforwardly build up the Yukawa sector for the quarks in the model basis using Table \ref{Tab:CompQuantumNumbers},
\begin{equation}
\begin{split}
\mathcal{L}_{Y} & \supset a_u \,\bar{Q}_L^{''1} u_R^{''3} + b_u \left[\bar{Q}_L^{''23} \phi_u \right]_{\bf{1_{++}}} u_R^{''1} + c_u \left[\bar{Q}_L^{''23} \phi_u \right]_{\bf{1_{-+}}} u_R^{''2} \\
&+ a_d \,\bar{Q}_L^{''1} d_R^{''3} + b_d \left[\bar{Q}_L^{''23} \phi_d \right]_{\bf{1_{++}}} d_R^{''1} + c_d \left[\bar{Q}_L^{''23} \phi_d \right]_{\bf{1_{-+}}} d_R^{''2}\,,
\end{split}
\end{equation}
where we have omitted Higgs fields and scale suppressions.  Using the vevs from \eqref{compvevs} and product rules from Appendix \ref{app:complicated}, we get the following Yukawa matrices: 
\begin{equation}
Y^{''\dagger}_u = v_u \begin{pmatrix}
0 & 0 & a_u/v_u \\
b_u & -c_u & 0 \\
-b_u e^{-i\pi/7} & -c_u e^{-i\pi/7} & 0
\end{pmatrix}, \ \ \ \ 
Y^{''\dagger}_d = v_d \begin{pmatrix}
0 & 0 & a_d/v_d \\
b_d & -c_d & 0 \\
-b_d & -c_d & 0  
\end{pmatrix}.
\end{equation}
Assembling these into their Hermitian combinations, one immediately finds 
\begin{align}
\nonumber
Y_u^{\prime \prime \dagger} Y_u^{\prime \prime} &= v_u^2\left(
\begin{array}{ccc}
\vert a_u \vert^2/v_u^2 & 0 & 0 \\
0 & |b_u|^2 + |c_u|^2  & e^{i\pi/7}(|c_u|^2 - |b_u|^2) \\
0 & -e^{-i\pi/7}(|b_u|^2 - |c_u|^2) & |b_u|^2 + |c_u|^2 
\end{array}
\right)\,, \\
Y_d^{\prime \prime \dagger} Y_d^{\prime \prime} &= v_d^2 \left(
\begin{array}{ccc}
\vert a_d \vert^2/v_d^2 & 0 & 0 \\
0 & |b_d|^2 + |c_d|^2  &  |b_d|^2 - |c_d|^2\\
0 & |b_d|^2 - |c_d|^2 & |b_d|^2 + |c_d|^2 
\end{array}
\right)\,,
\end{align}
which directly map to \eqref{eq:CompgenMass} with 
\begin{align}
\nonumber
\vert a_u \vert^2 \leftrightarrow \vert m_{u_3} \vert^2, \,\,\,\,\, \vert b_u \vert^2 \leftrightarrow \vert m_{u_2} \vert^2,  \,\,\,\,\, \vert c_u \vert^2 \leftrightarrow \vert m_{u_1} \vert^2 \,,
\end{align}
and analogous relations for the mapping of $Y_d$, up to VEV factors and multiplicative constants.  We have therefore reconstructed a successful top-down model, as it exhibits the required symmetry breaking in \eqref{eq:Compdiag} and recovers the CKM mixing prediction in \eqref{eq:CKMComp}.

\section{Application to Models of Flavourful Leptoquarks}
\label{sec:LEPTOQUARKS}

As a particularly relevant extension of the field content of the SM, we now apply our algorithm to a class of flavoured leptoquark models defined in \cite{deMedeirosVarzielas:2019lgb,Bernigaud:2019bfy}, which we will briefly review for completeness. Leptoquarks have been become popular in the recent literature due to their ability to resolve (potential) anomalies in heavy meson decay observables like $\mathcal{R}_{K^{\star}}$ observed by LHCb \cite{Aaij:2019wad,Aaij:2017vbb}, as well as other potentially anomalous measurements sensitive to muon physics (see e.g. \cite{Chatrchyan:2013bka,Aaij:2017vad,Aaboud:2018mst,Aaij:2020nrf}).  Here we allow the SM to be augmented by one of the following bosons, denoted the `scalar triplet,' `vector singlet,' and `vector triplet,' whose charge assignments under the SM gauge group are respectively given by (in the form $\mathcal{G}_{SM} \equiv SU(3)_C \times SU(2)_L \times U(1)_Y$),
\begin{equation}
\label{eq:leptoreps}
\Delta_{3} \sim \left({\bf{\bar{3}}}, {\bf{3}}, 1/3\right),\,\,\,\,\,\Delta_{1}^{\mu} \sim \left({\bf{3}}, {\bf{1}}, 2/3\right),\,\,\,\,\,\Delta_{3}^{\mu} \sim \left({\bf{3}}, {\bf{3}}, 2/3\right)\, .
\end{equation}
These leptoquarks are easily motivated in the UV by Grand Unified constructions, or in models with new gauge interactions (see e.g. \cite{Perez:2013osa, Biggio:2016wyy}), and all can successfully account for $\mathcal{R}_{K^{(*)}} < 1$ \cite{Hiller:2017bzc}. The SM-gauge invariant operators they source are given by
\begin{align}
\nonumber
\Delta_{3}&: \,\,\,\,\,&&\mathcal{L} \supset y_{3, ij}^{LL} \bar{Q}_{L}^{C\,i,a} \epsilon^{ab} (\tau^{k} \Delta_{3}^{k})^{bc} L_{L}^{j,c} + z_{3,ij}^{LL} \bar{Q}_{L}^{C\,i,a}\epsilon^{ab}((\tau^{k}\Delta_{3}^{k})^{\dagger})^{bc}Q_{L}^{j,c} + \text{h.c.} \\
\nonumber
\Delta_{1}^{\mu}&: \,\,\,\,\,&&\mathcal{L} \supset x_{1,ij}^{LL} \bar{Q}_{L}^{i,a} \gamma^{\mu} \Delta_{1,\mu} L_{L}^{j,a} + x_{1,ij}^{RR} \bar{d}^{i}_{R} \gamma^{\mu} \Delta_{1,\mu} E_{R}^{j} + x_{1,ij}^{\overline{RR}} \bar{u}_{R}^{i} \gamma^{\mu} \Delta_{1,\mu} \nu_{R}^{j} + \text{h.c.}\\
\label{eq:LLyukSU2}
\Delta_{3}^{\mu}&: \,\,\,\,\,&&\mathcal{L}  \supset  x_{3,ij}^{LL} \bar{Q}_{L}^{i,a} \gamma^{\mu} \left(\tau^{k} \Delta_{3,\mu}^{k}\right)^{ab} L_{L}^{j,b}  + \text{h.c.} 
\end{align}
with $\lbrace i,j \rbrace$ denoting flavour indices, $\lbrace a,b \rbrace$ denoting SU(2) indices, and $k = 1,2,3$ for the Pauli matrices.\footnote{The physics of leptoquarks is thoroughly reviewed in \cite{Dorsner:2016wpm}.}  Following \cite{Dorsner:2016wpm} and redefining the components of the scalar triplet state according to,\footnote{We will write the following equations explicitly for the scalar triplet, although analogous expressions are easily derived for the other two leptoquark states of \eqref{eq:leptoreps}.  Superscripts on the LHS denote electric charges.}
\begin{align}
\label{eq:combo1}
&\Delta_{3}^{4/3} = \left(\Delta_{3}^{1} - i \Delta_{3}^{2} \right)/\sqrt{2}, \,\,\,\,\,\,\,\,\,\,\,
&&\Delta_{3}^{-2/3} = \left(\Delta_{3}^{1} + i \Delta_{3}^{2} \right)/\sqrt{2}, \,\,\,\,\,\,\,\,\,\,\,
&&\Delta_{3}^{1/3} = \Delta_{3}^{3}  \, , 
\end{align}
contracting the SU(2) indices of \eqref{eq:LLyukSU2}, and ignoring the diquark operator of $\Delta_3$ (for simplicity, although RFS can control it--- see \cite{deMedeirosVarzielas:2019lgb}, and it can also be controlled with other symmetries \cite{Hambye:2017qix, Davighi:2020qqa}) , one then finds that the Yukawa/mass sector of the SM is enhanced to 
\begin{align}
\nonumber
\mathcal{L}_{mass} \,\,\, \supset \,\,\,&\frac{1}{2} \bar{\nu}^{c}_{L}\, m_{\nu}\, \nu_{L} + \bar{E}_{R}\, m_{l}\, l_{L} + \bar{d}_{R}\, m_{d}\, d_{L} + \bar{u}_{R}\, m_{u} \,u_{L}   \\
\nonumber
&+\, \bar{d}^{C}_{L} \,\lambda_{dl}\, l_{L} \, \Delta^{4/3}_{3} + \bar{d}^{C}_{L} \,\lambda_{d\nu}\, \nu_{L} \, \Delta^{1/3}_{3} + \bar{u}^{C}_{L} \,\lambda_{ul}\, l_{L} \, \Delta^{1/3}_{3} + \bar{u}^{C}_{L} \,\lambda_{u\nu}\, \nu_{L} \, \Delta^{-2/3}_{3} \\
\label{eq:LQYUKfull}
&+\, \text{h.c.}\,
\end{align}
with the novel leptoquark couplings $\lambda_{QL}$ normalized to the first term, $\lambda_{dl}$, which in the mass-eigenstate basis of the SM fermions is generically parameterized by
\begin{equation}
\label{eq:genericyuk}
-\sqrt{2}\,\left(U_{d}^{T}\,y_{3}^{LL}\,U_{l} \right) \equiv \lambda_{dl} =
\left(
\begin{array}{ccc}
\lambda_{de} & \lambda_{d\mu} & \lambda_{d\tau}  \\
\lambda_{se}  & \lambda_{s\mu}  & \lambda_{s\tau}     \\
\lambda_{be}  & \lambda_{b\mu}   & \lambda_{b\tau}  
\end{array}
\right) \, . 
\end{equation}
The other couplings in \eqref{eq:LQYUKfull} are related to $\lambda_{dl}$ via SU(2) relations, and are given by
\begin{align}
\label{eq:LQrelations}
\lambda_{d \nu} &= \frac{1}{\sqrt{2}} \lambda_{dl}\,U_{PMNS}, \,\,\,\,\,\, &&\lambda_{ul} = \frac{1}{\sqrt{2}} U^{\star}_{CKM}\,\lambda_{dl}, \,\,\,\, &&\lambda_{u \nu} = - U^{\star}_{CKM}\,\lambda_{dl}\,U_{PMNS}\,.
\end{align}

Given these new flavoured couplings, we defined multiple classes of simplified models based on the RFS formalism in \cite{deMedeirosVarzielas:2019lgb,Bernigaud:2019bfy} .  In particular, we assumed that the natural RFS of the SM (cf. \eqref{eq:SMRFS}) also hold in the new leptoquark terms of \eqref{eq:LQYUKfull}.  This allowed us to constrain the $\lambda_{QL}$ couplings via RFS invariances of the form
\begin{equation}
\exists\, \lbrace Q, L \rbrace\,, \,\,\,  T^{(T,\dagger)}_Q\,\lambda_{QL}\,T_L \overset{!}{=} \lambda_{QL}\,,
\end{equation}
with $Q$, $L$ respectively representing arbitrary quark ($d$, $u$) and lepton ($l$, $\nu$) families, and the transposed `$T$' (daggered `$\dagger$') $T_Q$ corresponding to scalar (vector) leptoquark(s).  Critically, different phase relationships amongst RFS generators $T_{Q,L}$ correspond to different textures in $\lambda_{QL}$, and the extent to which free parameters remain in \eqref{eq:genericyuk}-\eqref{eq:LQrelations} is a function of the amount of symmetry present in any given term.  Precision flavour data from (e.g.) $B-\overline{B}$ mixing, lepton-flavour-violating (LFV) observabes like $\mu \rightarrow e \gamma$, and the anomalous $\mathcal{R}$ ratios also constrain the viable textures (and hence also the phenomenologically viable RFS relationships) in \eqref{eq:genericyuk}. 

For example, in \cite{deMedeirosVarzielas:2019lgb} we insisted that RFS hold in \emph{all} lepton and quark sectors of the SM and leptoquark couplings, and this led to only $\mathcal{O}(10)$ viable textures for $\lambda_{dl}$ with only a single real parameter, once all experimental and symmetry constraints were made.\footnote{We also insisted that at least two generations be distinguished by the eigenvalues of $T_{Q,L}$, so that the action of the RFS was not trivial in a given family sector.}  Then, in \cite{Bernigaud:2019bfy}, we relaxed the symmetry assumptions and enforced an RFS invariance in some or all of the SM mass terms, but \emph{only} in the $d-l$ leptoquark coupling, where either or both $T_{d,l}$ were allowed to act; in this symmetry environment, invariance in $\lambda_{d\nu, ul, u\nu}$ is inherited via SU(2) relationships as in \eqref{eq:LQrelations}.  These two types of simplified models were labeled `SE1' and `SE2' respectively, with the former likely requiring more intricate model building to account for the fact the RFS distinguishes members of SU(2) doublets in individual leptoquark terms after EWSB.  The SE2 models, on the other hand, represent highly natural relaxations of the SE1 constructions, and can easily be realized with single flavon EFTs as per our algorithm in Section \ref{sec:THEORY}, which we will now show.

\subsection*{The Leptoflavour Basis}
Viable NADS that can realize the SE2 symmetry predictions for $U_{PMNS,CKM}$ and $\lambda_{QL}$ must be uncovered in order to apply our algorithm (cf. Section \ref{sec:SCAN}), and to that end a {\tt{GAP}} scan was performed in \cite{Bernigaud:2019bfy}. We performed that scan in the `leptoflavour basis', where information about all relevant physical mixings could be extracted, and which we now review.  

Recall that, in the mass-eigenstate basis, the RFS generators (and therefore $\mathcal{G_{F,L,Q}}$) do not know about $U_{PMNS,CKM}$, even in the case of only SM field content.  However, we can go to a special form of the `flavour basis' (cf. \eqref{eq:SMfermionflav}) by doing no change of basis (or trivial change of basis with the identity matrix) on the LH $d$, $l$, and a change of basis via the CKM for the LH $u$ and the PMNS for the LH $\nu$.  With leptoquarks present, we simultaneously want to encode the additional information present in $\lambda_{QL}$ together with the SM mixing matrices, so we choose the leptoflavour basis to be the one where the SM mixing is all in the mixing matrices, and therefore the charged current should be diagonal --- similarly to what we have in the flavour basis --- but additionally, we choose to have $\lambda_{dl}$ diagonal. To diagonalize $\lambda_{dl}$, we require a further non-trivial change of basis in the LH $d$, $l$, which must be cancelled with rotations in the $u$ and $\nu$ respectively, to have the CC diagonal. So, starting from the mass basis (where $\lambda_{dl}$ is defined as in \eqref{eq:genericyuk}), $d$ changes by $\Lambda_d$, $l$ by $\Lambda_l$ and $u$ changes by both CKM and by $\Lambda_d$ (canceling the presence of $\Lambda_d$ in the CC and making it diagonal), while $\nu$ changes by both PMNS and by $\Lambda_l$ (canceling the presence of the $\Lambda_l$ in the CC, and making it diagonal). Finally, the case without leptoquarks appears correctly as the limit with $\Lambda_d = \Lambda_l = 1$.

To see this explicitly, we apply the following `leptoflavour' basis transformations on the mass eigenstates:
\begin{align}
\nonumber
l_L &\rightarrow \Lambda_{l}^{\dagger} l'_L,  &&d_L \rightarrow \Lambda_{d}^\dagger d'_L, &&\nu_L \rightarrow U_{PMNS}^\dagger \Lambda_{l}^\dagger \nu'_L,  &&u_L \rightarrow U_{CKM} \Lambda_{d}^\dagger u'_L, \\
\label{eq:leptoflavourbasistransform}
E_{R} &\rightarrow \Lambda_{E}^{\dagger} E_{R}^{'}, \,\,\,\,\,\,\,\, &&d_R \rightarrow \Lambda_{D}^\dagger d'_R, &&\nu_R \rightarrow  \Lambda_{R}^\dagger \nu'_R, &&u_R \rightarrow  \Lambda_{U}^\dagger u'_R.
\end{align}
This yielded the leptoflavour basis Lagrangian,
\begin{align}
\nonumber
\mathcal{L} &\supset  \frac{g}{\sqrt{2}}\bar{l}'_L  \gamma^\mu \nu'_L W_{\mu}^- + \frac{g}{\sqrt{2}}\bar{d}'_L \gamma^\mu u'_L W_{\mu}^- \\
\nonumber
&+ \frac{1}{2}\bar{\nu}_{L}^{'c}\Lambda_{l}^* U_{PMNS}^* m_\nu U_{PMNS}^\dagger \Lambda_{l}^\dagger \nu'_L +\bar{E}^{'}_R \Lambda_{E} m_l \Lambda_{l}^\dagger l'_L + \bar{d}_{R}^{'}\Lambda_{D}  m_d \Lambda_{d}^\dagger d'_L + \bar{u}^{'}_R \Lambda_{U}  m_u U_{CKM}\Lambda_{d}^\dagger u'_L \\
\nonumber
& + \frac{1}{\sqrt{2}}\bar{d}_{L}^{'c} \Lambda_{d}^* \lambda_{dl} \Lambda_{l}^\dagger\nu'_L \Delta_{3}^{1/3} + \bar{d}_{L}^{'c} \Lambda_{d}^* \lambda_{dl} \Lambda_{l}^\dagger l'_L \Delta_{3}^{4/3}  + \bar{u}_{L}^{'c} \Lambda_{d}^* \lambda_{dl} \Lambda_{l}^\dagger\nu'_L \Delta_{3}^{-2/3} + \frac{1}{\sqrt{2}}\bar{u}_{L}^{'c} \Lambda_{d}^* \lambda_{dl} \Lambda_{l}^\dagger l'_L \Delta_{3}^{1/3} \\
\label{eq:Flavour_basis_Lagrangian}
&+\text{h.c.},
\end{align}
which is invariant under the following LH RFS generators:
\begin{equation}
\label{Eq:LGenerators_Flavour_basis}
T_{l}^{'} = \Lambda_l T_l \Lambda_{l}^\dagger, \,\,\,\,\,T_{\nu}^{'} = \Lambda_{l} U_{PMNS} T_\nu U_{PMNS}^\dagger \Lambda_{l}^\dagger,\,\,\,\,\, T_{d}^{'}=\Lambda_d T_d \Lambda_{d}^\dagger, \,\,\,\,\, T_{u}^{'} = \Lambda_d U_{CKM}^\dagger T_{u} U_{CKM} \Lambda_{d}^\dagger,
\end{equation}
and the following RH RFS generators:
\begin{equation}
\label{eq:RHgens}
T_{E}^{'}=\Lambda_{E} T_l \Lambda_{E}^{\dagger},\ \ \ \ T_{R}^{'} = \Lambda_{R} T_\nu \Lambda_{R}^{\dagger}, \ \ \ \ T_{D}^{'} = \Lambda_{D} T_d \Lambda_{D}^{\dagger}, \ \ \ \ T_{U}^{'} = \Lambda_{U} T_u \Lambda_{U}^{\dagger},
\end{equation}
where $T_R'$ holds only in the case of Dirac neutrinos.  As mentioned above, in the limit where $\Lambda_{l,d} \rightarrow \mathbb{1}$, \eqref{Eq:LGenerators_Flavour_basis} reduces to the SM-only flavour-basis generating set!  We also point out that, in the absence of a RH leptoquark coupling as present in (e.g.) the vector singlet case, one has some freedom to choose the RH $\Lambda_{E,R,D,U}$ transformations since their shapes are not dictated by the requirement of diagonalizing a particular coupling (we can always form totally LH combinations of the SM mass matrices, e.g.). Given the leptoflavour basis, bottom-up scans as described in Section \ref{sec:SCAN} were then performed in \cite{Bernigaud:2019bfy} with \eqref{Eq:LGenerators_Flavour_basis}, so that parent family groups were closed according to \eqref{eq:GFall}-\eqref{eq:GQ}.  As it turns out, many NADS were discovered, including members of the popular $S_N$, $A_N$, $\Delta(3N^2)$, $\Delta(6N^2)$, $\Sigma(3N^2)$, $\Sigma(3N^3)$, and $D_N$ finite group series.  

\begin{table}[t]
\renewcommand{\arraystretch}{1.5}
\centering
\begin{tabular}{|c||c|c|c|}
\hline
\multicolumn{4}{|c|}{\textbf{Basis-Dependent RFS Quantities}}\\
\hline
\hline 
$\text{Basis}$ & $\text{Transform $\nu_L \rightarrow $}$ & $\text{Mass Matrix}$ & $\text{RFS Generator}$\\
\hline
\hline
$\text{Mass}$ & $\nu_L$ & $m_\nu$ & $T_\nu$ \\
\hline
$\text{Flavour}$ & $U_\nu^{\dagger} \nu_L$ &  $U_\nu^\star \, m_\nu \,U_\nu^\dagger$  & $U_\nu\, T_\nu \, U_\nu^{\dagger}$\\
\hline
$\text{Leptoflavour}$ & $U_{PMNS}^\dagger \Lambda_{l}^\dagger \nu'_L$ & $\Lambda_{l}^* U_{PMNS}^* m_\nu U_{PMNS}^\dagger \Lambda_{l}^\dagger$ &   $\Lambda_{l} U_{PMNS} T_\nu U_{PMNS}^\dagger \Lambda_{l}^\dagger$\\
\hline
$\text{Model}$ & $U_{PMNS}^\dagger \Lambda_{l}^\dagger P \nu^{\prime\prime}_L$ & $P^T\Lambda_{l}^* U_{PMNS}^* m_\nu U_{PMNS}^\dagger \Lambda_{l}^\dagger P$ & $P^\dagger \Lambda_{l} U_{PMNS} T_\nu U_{PMNS}^\dagger \Lambda_{l}^\dagger P$\\
\hline
\end{tabular}
\caption{The RFS generators associated with a given basis change away from the original mass-eigenstate $\nu_L$.}
\label{tab:GenChange}
\end{table}

As a final preparation for the reconstruction of the RFS-invariant Lagrangian in these extended leptoquark scenarios, we allow for the possibility that an additional basis change will be amenable for manipulating the group product rules of the NADS discovered in \cite{Bernigaud:2019bfy}.  Hence we rotate via a generic matrix $P$ (which can be set to the identity matrix in the event it is unnecessary), 
\begin{equation}
a^{\prime} \rightarrow P a^{\prime\prime} ,
\end{equation}
and so the effective mass terms are now given by
\begin{align}
\nonumber
\mathcal{L} &\supset \frac{1}{2}\bar{\nu}_{L}^{''c}\,\underbrace{P^T\Lambda_{l}^* U_{PMNS}^* m_\nu U_{PMNS}^\dagger \Lambda_{l}^\dagger P\,}_{m^{''}_{\nu}} \nu^{''}_L +\bar{u}^{''}_R \, \underbrace{P^{\dagger} \Lambda_{U}  m_u U_{CKM}\Lambda_{d}^\dagger P}_{m^{''}_u} \, u^{''}_L \, \\
\label{eq:upnuNEW}
&+ \bar{E}^{''}_R \underbrace{P^\dagger \Lambda_{E} m_l \Lambda_{l}^\dagger P}_{m_l^{''}} l^{''}_L + \bar{d}_{R}^{''}\underbrace{P^{\dagger}\Lambda_{D}  m_d \Lambda_{d}^\dagger P}_{m_d^{''}} d^{''}_L ,
\end{align}
where we have assumed that $A_R$ also transforms with $P$.  The mass matrices in this basis are labeled by $m^{''}$, and are clearly non-diagonal.  The remaining leptoquark terms of \eqref{eq:LQYUKfull} are similarly given in this basis by
\begin{align}
\nonumber
\mathcal{L} &\supset  \frac{1}{\sqrt{2}}\bar{d}_{L}^{\prime\prime c}\underbrace{P^T \Lambda_{d}^* \lambda_{dl} \Lambda_{l}^\dagger P}_{\lambda_{d\nu}^{\prime\prime}} \nu^{\prime\prime}_L \Delta_{3}^{1/3} + \bar{d}_{L}^{\prime\prime c} \underbrace{P^T \Lambda_{d}^* \lambda_{dl} \Lambda_{l}^\dagger P}_{\lambda^{\prime \prime}_{dl}} l^{\prime\prime}_L \Delta_{3}^{4/3}  \\
\label{eq:leptoRFSfinal}
&+ \bar{u}_{L}^{\prime\prime c} \underbrace{P^T \Lambda_{d}^* \lambda_{dl} \Lambda_{l}^\dagger P}_{\lambda^{\prime \prime}_{u\nu}} \nu^{\prime\prime}_L \Delta_{3}^{-2/3} + \frac{1}{\sqrt{2}}\bar{u}_{L}^{\prime\prime c} \underbrace{P^T\Lambda_{d}^* \lambda_{dl} \Lambda_{l}^\dagger P}_{\lambda^{\prime \prime}_{ul}} l^{\prime\prime}_L \Delta_{3}^{1/3}\,.
\end{align}
These operators already reveal the natural form of their respective RFS generators.  A summary of all basis changes on (e.g.) the neutrino field and associated changes in $m_\nu$ and $T_\nu$ are given in Table \ref{tab:GenChange}, tracking all the way from the mass-eigenstate basis to the model basis of \eqref{eq:upnuNEW}-\eqref{eq:leptoRFSfinal}.

Hence, given a specific NADS, its RFS, and the associated predictions for $U_{CKM,PMNS}$ and $\lambda_{dl}$, one can use \eqref{eq:upnuNEW}-\eqref{eq:leptoRFSfinal} to reconstruct the UV EFT.  We will now consider two such models, one based on the $\Delta(96)$ group and one based on the $D_{15}$ member of the Dihedral series $D_N$.  All of the relevant bottom-up information for these groups is given in Table \ref{tab:NADSresult}, and the parameters $x_{e,\mu}$ are defined in the following textures:
\begin{equation}
\label{eq:yukeisolation0}
\lambda^{[e0]}_{dl} = 
\lambda_{be} \left(
\begin{array}{ccc}
0 & 0 & 0  \\
x_e  & 0 & 0    \\
1  & 0  & 0 
\end{array}
\right), \,\,\,\,\,
\lambda^{[\mu0]}_{dl} = 
 \lambda_{b\mu} \left(
\begin{array}{ccc}
0 & 0 & 0  \\
0 & x_{\mu} & 0  \\
0 & 1 & 0 
\end{array}
\right), \,\,\,\,\, \text{with} \,\,\,\, x_{X} = \frac{\lambda_{sX}}{\lambda_{bX}}\,,
\end{equation}
which are the consequence of special relationships amongst RFS generators --- $\lambda^{[e0]}_{dl}$ corresponds to $-\alpha_{l} = \beta_{d} = \gamma_{d}$ while $\lambda^{[\mu0]}_{dl}$ corresponds to $-\beta_{l} = \beta_{d} = \gamma_{d}$.\footnote{These are the relationships for the scalar triplet, while for the vector singlet and triplet the minus signs do not appear.  See \cite{Bernigaud:2019bfy} for details, where other couplings are also controlled.}  By construction these couplings are diagonalized with $\Lambda_{d,l}$ in the combination $\Lambda_d^\star \lambda_{dl} \Lambda_l^\dagger$, with 
\begin{equation}
\label{eq:Lambdas}
\Lambda_d(x_{e/\mu}) = \left(
\begin{array}{ccc}
0 & \frac{x_{e,\mu}}{\sqrt{1+x_{e,\mu}^2}} & \frac{1}{\sqrt{1+x_{e,\mu}^2}}\\
0 & -\frac{1}{\sqrt{1+x_{e,\mu}^2}} & \frac{1}{\sqrt{1+1/x_{e,\mu}^2}}\\
1 & 0 & 0
\end{array}
\right)\,,\,\,\,\,\,\,\,\,\,\,
\Lambda_l^e = \left(
\begin{array}{ccc}
1 & 0 & 0 \\
0 & 0  & 1 \\
0 & 1 & 0
\end{array}
\right)\,, \,\,\,\,\,\,\,\,\,\,\,\,
\Lambda_l^\mu = \left(
\begin{array}{ccc}
0 & 1 & 0\\
0 & 0 & 1\\
1 & 0 & 0
\end{array}
\right)\,.
\end{equation}
The quark matrix $\Lambda_d$ in \eqref{eq:Lambdas} left-diagonalizes both patterns in \eqref{eq:yukeisolation0}, whereas $\Lambda_l^e$ right-diagonalizes $\lambda_{dl}^{[e0]}$ and $\Lambda_l^\mu$ right-diagonalizes $\lambda_{dl}^{[\mu0]}$.

On the other hand, the parameters $t_{\theta_{\mu\tau}} \equiv \tan \theta_{\mu\tau}$ and $\theta_C$ are defined in the following LO PMNS and CKM textures:
\begin{align}
\label{eq:mutauPMNS}
U_{PMNS} \simeq U_{\mu \tau} &\equiv \frac{1}{\sqrt{2}}
\left(
\begin{array}{ccc}
\sqrt{2} \cos \theta_{\mu\tau} & \sqrt{2} \sin \theta_{\mu\tau} & 0  \\
-\sin \theta_{\mu\tau} &  \cos \theta_{\mu\tau} & 1   \\
\sin \theta_{\mu\tau}   & - \cos \theta_{\mu\tau}  & 1
\end{array}
\right) + \mathcal{O}\left(\theta_{13}^{l}\right),
\\
\label{eq:Cabibbo}
U_{CKM} \simeq U_{C} &\equiv 
\left(
\begin{array}{ccc}
\cos \theta_{C} & \sin \theta_{C} & 0  \\
- \sin \theta_{C} & \cos \theta_{C} & 0    \\
0   & 0 & 1
\end{array}
\right) + \mathcal{O}\left( \theta_{C}^{2}, \theta_{C}^{3} \right),
\end{align}
which were specified as the SM mixing to be recovered in \cite{Bernigaud:2019bfy}.\footnote{After all, ours is a bottom-up approach, and therefore closing finite groups via \eqref{eq:GFall}-\eqref{eq:GQ} requires that some textures for $U_{PMNS,CKM}$ are fed to the algorithm.}
While the exact forms of \eqref{eq:mutauPMNS} and \eqref{eq:Cabibbo} are excluded by current global fits to experiment \cite{Esteban:2018azc}, they nevertheless provide excellent approximations to the data.  The $\mu-\tau$ symmetric matrix in \eqref{eq:mutauPMNS}, for example, reproduces global fits to the PMNS matrix up to corrections on the order of the smallest `reactor angle,' $\theta^l_{13}$, and its free parameter $\theta_{\mu\tau}$ can be fit to many well-studied textures like the tri-bimaximal \cite{Harrison:2002er}, bi-maximal \cite{Fukugita:1998vn}, golden ratio \cite{Datta:2003qg,Adulpravitchai:2009bg}, and hexagonal forms \cite{Giunti:2002sr,Albright:2010ap}:
\begin{equation}
\label{eq:mt2}
U_{\mu \tau}\left(\theta_{\mu\tau}\right) \rightarrow \begin{cases}
								U_{TBM}  &\rightleftarrows \tan\theta_{\mu\tau} = \frac{1}{\sqrt{2}}
								\\
								U_{BM}  &\rightleftarrows \tan \theta_{\mu\tau}  = 1 \,\,\, \text{or} \,\,\,\theta_{\mu\tau} =\frac{\pi}{4}
								\\
								U_{GR_{1}} &\rightleftarrows \tan\theta_{\mu\tau}= \frac{2}{(1+\sqrt{5})}
								\\
								U_{GR_{2}} &\rightleftarrows \theta_{\mu\tau} =\frac{\pi}{5}
								\\
								U_{HM} &\rightleftarrows \tan\theta_{\mu\tau} = \frac{1}{\sqrt{3}}\,\,\, \text{or} \,\,\,\theta_{\mu\tau} = \frac{\pi}{6} \,.
								\end{cases}
\end{equation}
Furthermore, corrections to \eqref{eq:mutauPMNS} can naturally be realized by higher-order terms in the EFT expansion beyond \eqref{eq:fullayuka}, which can softly break the RFS embedded in the LO contribution, or also through renormalization group flow (RGE) \cite{Olechowski:1990bh,Ross:2007az,Varzielas:2008jm,Chiu:2016qra,Casas:1999tg,Casas:1999ac,Chankowski:1999xc,Antusch:2003kp} between the scale at which $\mathcal{G_F}$ is broken and the IR, where global fits are performed.  

Similarly, \eqref{eq:Cabibbo} provides an excellent description of the dominant Cabibbo mixing of the CKM matrix.  Unlike the PMNS, the CKM is extremely hierarchical, with mixings in the (2,3) and (1,3) sectors suppressed by one to two orders of magnitude with respect to the Cabibbo sector.  This suppression again hints at further contributions to \eqref{eq:Cabibbo} from higher-order terms in \eqref{eq:fullayuka} and/or RGE corrections.  

Of course, precisely calculating the corrections expected to \eqref{eq:mutauPMNS}-\eqref{eq:Cabibbo} depends on the complete UV flavour model, including not only the full field and symmetry content, but also the presence or lack thereof of supersymmetry.  Specifying this is well beyond the scope of our present paper, and so we consider \eqref{eq:mutauPMNS}-\eqref{eq:Cabibbo} sufficiently accurate to develop our approach to reconstructing effective Lagrangians from RFS.

\begin{table}[t]
\renewcommand{\arraystretch}{1.5}
\makebox[\textwidth]{
\begin{tabular}{|c||c|c|c|c||c|c|}
\hline
\multicolumn{7}{|c|}{\textbf{Finite Groups, RFS, and Mixing Predictions}}\\
\hline
\hline 
 $\lbrace x_{e/\mu}, t_{\theta_{\mu\tau}}, \theta_C  \rbrace$ & $T^{ii}_{d}$  & $T^{ii}_{l}$  &  $T^{ii}_{u}$ &  $T^{ii}_{\nu}$ & GAP-ID & $\mathcal{G_F}$  \\
\hline
\hline
 $\lbrace \text{N.A.},\frac{1}{\sqrt{2}},\text{N.A.}\rbrace$ & \text{N.A.} & [1,$\omega_3$,$\omega_3^2$]  & \text{N.A.} & [$\omega_4$,1,-$\omega_4$] & [96, 64] & $\Delta(96)$ \\
\hline
$\lbrace1,\frac{1}{\sqrt{3}},\frac{\pi}{15}\rbrace$ & [1,-1,-1] & [1,-1,-1] & [1,-1,-1]& [-1,1,-1]&  [30,3]  & $D_{15}$ \\
\hline
$\lbrace1,1,\frac{\pi}{14}\rbrace$ & [1,-1,-1] & [1,-1,-1] & [1,-1,-1] & [$\omega_4$,-$\omega_4$,1] & [56,7] & $\left(\mathbb{Z}_{14} \times \mathbb{Z}_2 \right) \rtimes \mathbb{Z}_{2} $ \\
\hline
\end{tabular}}
\caption{The scan results from \cite{Bernigaud:2019bfy} that we will use to reconstruct models in Sections \ref{sec:D96}-\ref{sec:D15}.  Note that we have relabeled $D_{30} \rightarrow D_{15}$ from \cite{Bernigaud:2019bfy} in order to reflect the conventions of \cite{Ishimori:2010au}.}
\label{tab:NADSresult}
\end{table}

\subsection{$\Delta(96)$ Model for $U_{PMNS}$ and Leptoquarks}
\label{sec:D96}
As a first example incorporating leptoquarks we construct a $\Delta(96)$ flavour model from the scan results in \cite{Bernigaud:2019bfy}, which are repeated in the first row of Table \ref{tab:NADSresult}. This model predicts tri-bimaximal mixing $U_{PMNS} = U_{\mu\tau}(\arctan 1/\sqrt{2})$ and the electron isolation pattern $\lambda_{dl}^{[e0]}$ for the leptoquark coupling. While the muon isolation pattern is phenomenologicaly preferred over the electron one, as it explains further the different muon anomalous observables while the electron isolation pattern could only explain the deviation in $R_K^{(*)}$, we focus here on the electron isolation as this rather simple group can predict it unambiguously. We highlight the fact that other symmetries, such $A_4$ or $\Delta (75)$, obtained using the procedure described in \cite{Bernigaud:2019bfy} are capable of reproducing the muon isolation pattern.

The symmetry breaking to RFS is illustrated in
\begin{equation}
\label{eq:D96diag}
\mathcal{G_{L}} \cong \Delta(96)   \rightarrow	 \begin{cases} 
							
				            		 \mathcal{G_{\nu}} \cong \mathbb{Z}_4
					     		 \\
					     		 \mathcal{G_{\text{l}}} \, \cong \mathbb{Z}_3 \,\,.
							
						\end{cases} \\
\end{equation}
As evident from the fact that  $T_\nu$ generates $  \mathcal{G_{\nu}} \cong \mathbb{Z}_4$, this model features Dirac rather than Majorana neutrino masses. In Table \ref{tab:NADSresult} one can read off the specific charges of the mass-eigenstate RFS generators $T_l$ and $T_\nu$. We note their basis-independent Traces are respectively $0$ and $1$, which will soon help us identify the conjugacy class to which they belong within $\Delta(96)$. 

Table \ref{tab:NADSresult} gives all the information required to move to the leptoflavour basis, where the RFS generators take the forms
\begin{equation}
T_l^{'} = \left(
\begin{array}{ccc}
1 & 0 & 0 \\
0 & \omega_3  & 0 \\
0 & 0 & \omega_3^2
\end{array}
\right)\,,\,\,\,\,\,\,\,\,\,\,
T_\nu^{'} = \frac{1}{3} \left(
\begin{array}{ccc}
1+2 i & -1+i & 1-i \\
-1+i & 1-i & -1-2i \\
1-i & -1-2i & 1-i
\end{array}
\right)\,,
\end{equation}
where $\omega_ 4=e^{i 2 \pi/4}=i$.
We want to identify these generators with group elements of $\Delta(96)$, and to do so we use the catalogue in \cite{Ding:2012xx}, repeating the relevant product rules in Appendix \ref{app:D96}.  As before, we find it convenient to perform a $P$ transformation on the leptoflavour basis, so that we go to a basis where the combination of $\Delta(96)$ generators $\bf{a_{3_1}^2 c_{3_1} d_{3_1}}$ for the $\bf{3_1}$ representation, found in \cite{Ding:2012xx}, is diagonal. We note that we use the same naming of the generators as in \cite{Ding:2012xx}, only differentiating them with the boldface to further avoid confusion with our naming for the coefficients (in this and in other sections). The $P$ matrix we use is
\begin{equation}
P = \left(
\begin{array}{ccc}
0 & 0 & 1 \\
0 & -1  & 0 \\
1 & 0 & 0
\end{array}
\right)\,,
\end{equation}
which leads to
\begin{equation}
\label{eq:TgensD96}
T_l^{''} 
= \left(
\begin{array}{ccc}
\omega_3 & 0 & 0 \\
0 &  \omega_3^2 & 0 \\
0 & 0 & 1
\end{array}
\right)\,,\,\,\,\,\,\,\,\,\,\,
T_\nu^{''}  
= 
\frac{1}{3} \left(
\begin{array}{ccc}
1-i & 1+2i & 1-i \\
1+2 i & 1-i & 1-i \\
1-i & 1-i & 1+2i
\end{array}
\right)\,.
\end{equation}
With this change of basis, we are able to match $T_l^{''}$ with the diagonal ${\bf{a^2 c d}}$ element of the $\bf{\bar{3}_1}$ representation of $\Delta(96)$, the conjugate representation to $\bf{3_1}$, which as expected has zero trace and lies within conjugacy class $\mathcal{C}_6$ \cite{Ding:2012xx}. According to the character of $T_\nu^{''}$, it could be within $\mathcal{C}_5$ or $\mathcal{C}_9$ in the same $\bf{\bar{3}_1}$ representation, and indeed we found it to match the element $\bf{a^2 b c^2 d^3}$.  With this information we introduce two flavons $\phi_{l,\nu}$, for which we use \eqref{eq:TgensD96} to derive candidate VEV in the triplet directions
\begin{equation}
\label{eq:D96vev}
\langle \phi_l \rangle =  \left(
\begin{array}{c}
0 \\
0 \\
v_l
\end{array}
\right)\,,\,\,\,\,\,\,\,\,\,\, \langle \phi_\nu \rangle =  \left(
\begin{array}{c}
v_\nu\\
v_\nu\\
v_\nu
\end{array}
\right)\,,
\end{equation}
which are invariant under $T_l^{''}$ for $\bf{\bar{3}_1}$ (and for $\bf{\bar{3}'_1}$) and $T_\nu^{''}$ for $\bf{\bar{3}_1}$, respectively.
Finally, the RFS-invariant mass combinations in this basis are given by
\begin{align}
\nonumber
m_{\nu}^{'' \dagger}m_{\nu}^{''}& \overset{!}{=}
\frac{1}{3} \left( 
\begin{array}{ccc}
\frac{1}{2} \left(m_{\nu_1}^2 + 2 m_{\nu_2}^2 + 3 m_{\nu_3}^2 \right) & \frac{1}{2} \left(m_{\nu_1}^2 + 2 m_{\nu_2}^2 - 3 m_{\nu_3}^2 \right) & \left(m_{\nu_2}^2 -  m_{\nu_1}^2 \right) \\
 \frac{1}{2} \left(m_{\nu_1}^2 + 2 m_{\nu_2}^2 - 3 m_{\nu_3}^2 \right)  &  \frac{1}{2} \left(m_{\nu_1}^2 + 2 m_{\nu_2}^2 + 3 m_{\nu_3}^2 \right)  &  \left(m_{\nu_2}^2 -  m_{\nu_1}^2 \right)\\
\left(m_{\nu_2}^2 -  m_{\nu_1}^2 \right)  & \left(m_{\nu_2}^2 -  m_{\nu_1}^2 \right) & \left(2m_{\nu_1}^2 -  m_{\nu_2}^2 \right)
\end{array}
\right), \\
\label{eq:genMassD96}
m_{l}^{''\dagger}m_{l}^{''}&\overset{!}{=}
\left( 
\begin{array}{ccc}
m_{l_2}^2 & 0 & 0 \\
0 & m_{l_3}^2 & 0 \\
0 & 0 & m_{l_1}^2 
\end{array}
\right),
\end{align}
which we will build below.  Note that there are no mixing ambiguities associated to these matrices.

\begin{table}[t]
    \centering
    \renewcommand{\arraystretch}{1.5}
    \begin{tabular}{|c||c|c|c|c|c|c|c|c|c|}
    \hline
  \  & $\bar{L}^{''}_L$ &  $(E_{R}^{''1}, E^{''2}_{R})$ & $E^{''3}_{R}$ & $(\nu_{R}^{''1},\nu_{R}^{''2})$ & $\nu_{R}^{''3}$ & $\phi_\nu$ & $\phi_{\nu2}$ & $\phi_l$ & $\phi_{l2}$ \\
  \hline
  \hline
  $\Delta(96)$ & $\bf{3_1}$ & $\bf{2}$ & $\bf{1}$ & $\bf{2}$ & $\bf{1}$ & $\bf{\bar{3}_1}$ & $\bf{\bar{3}_1}$  & $\bf{\bar{3}_1}$ & $\bf{\bar{3}_1}^\prime$ \\
  \hline
\end{tabular}    
\caption{Relevant field and $\Delta(96)$ symmetry content.}
    \label{Tab:D96QuantumNumbers}
\end{table}

\subsubsection*{The Lepton Sector}
We now build the model by assigning $L_L$ as a $\bf{\bar{3}_1}$, and the RH leptons as combinations of a $\Delta(96)$ trivial singlet and a doublet, which we designate as $\bf{1}+\bf{2}$ e.g. $E_{R}^{''3} \sim \bf{1}$, $E_{R}^{''12} \equiv \left(E_R^{''1}, E_R^{''2} \right) \sim \bf{2}$ and similarly for $\nu_{R}^{''3} \sim \bf{1}$ and $\nu_{R}^{''12} \sim \bf{2}$.
In this case the $\Delta(96)$ invariant Yukawa terms for charged leptons and for neutrinos are very similar, of the type 
\begin{equation}
\mathcal{L}_Y \supset a_{\nu,e} \, \left[ \bar{L}^{''}_L \phi_f\right]_{\bf{1}} f^{''3}_{R} + b_{\nu,e} \left[ \bar{L}^{''}_L \phi_f\right]_{\bf{2}} f^{''12}_{R} \,,
\end{equation}
where $f$ stands for either charged leptons or neutrinos and the Higgs field is omitted for simplicity.
For the neutrino sector we find that the $(1,1,1)$ direction gives rise to  
\begin{equation}
Y_\nu^{''\dagger} = 
a_\nu v_\nu \left(
\begin{array}{ccc}
0 & 0 & 1 \\
0 & 0 & 1 \\
0 & 0 & 1
\end{array}
\right)
+
b_\nu v_\nu \left(
\begin{array}{ccc}
\omega_3 & 1 & 0 \\
\omega_3 & 1 & 0 \\
\omega_3 & 1 & 0
\end{array}
\right)\,,
\end{equation}
which combines into
\begin{equation}
Y_\nu^{''\dagger} Y_\nu^{''} = 
(2 |b_\nu|^2 + |a_\nu |^2) \,v_\nu^2 \left(
\begin{array}{ccc}
1 & 1 & 1 \\
1 & 1 & 1 \\
1 & 1 & 1
\end{array}
\right)\,.
\end{equation}
A simplified version of this could be obtained through a shaping symmetry removing the coupling either to $\nu_R^{12}$ ($b_\nu=0$) or to $\nu_R^3$ ($a_\nu=0$). This matrix embeds the correct PMNS matrix predicted by the RFS framework, but with two massless neutrinos. Given that the charged lepton invariants are very similar, we can quickly construct the respective Yukawa matrix for this sector as well:
\begin{equation}
Y_l^{''\dagger} = 
a_e v_l \left(
\begin{array}{ccc}
0 & 0 & 0 \\
0 & 0 & 0 \\
0 & 0 & 1
\end{array}
\right)
+
b_e v_l \left(
\begin{array}{ccc}
0 & 1 & 0 \\
\omega_3 & 0 & 0 \\
0 & 0 & 0
\end{array}
\right)\,,
\end{equation}
leading to 
\begin{equation}
\label{eq:D96leptonA}
Y_l^{''\dagger} Y_l^{''}  = v_l^2
\left(
\begin{array}{ccc}
|b_e|^2 & 0 & 0 \\
0 & |b_e|^2 & 0 \\
0 & 0 & |a_e|^2
\end{array}
\right)\,.
\end{equation}
Hence the invariant operators give rise to a diagonal Yukawa coupling but with two degenerate charged lepton masses, which is clearly unphysical.

In order to make the model realistic, we first note that the directions  $\langle \phi_\nu \rangle \sim \lbrace(1,1,1)$, $(-1,1,0),(-1,-1,2)\rbrace^T$ are the eigenvectors of $T_\nu^{''}$ with eigenvalues $\hat{e}_\nu = \lbrace 1,-i, i \rbrace$, respectively.  While we initially selected the first eigensystem in \eqref{eq:D96vev} with $\hat{e}_\nu = 1$, according to \eqref{eq:vevconstrain} we are free to choose any of them, noting that while this doesn't actually preserve $T_\nu^{''}$ as a residual symmetry, the resulting mass matrices will still lead to a successful $Y_\nu^{''\dagger} Y_\nu^{''}$ in the sense that we obtain $U_{PMNS} = U_{\mu\tau}(\arctan 1/\sqrt{2})$ as intended. More details on this type of situation can be found in Appendix \ref{sec:APPB}. Taking either $(-1,1,0)$ or $(-1,-1,2)$ for an additional flavon $\phi_{\nu2}$'s orientation allows one to generate further non-zero masses in $m_\nu^{''}$. 

At the same time, for the charged leptons, it is possible to break the mass degeneracy  by having an additional triplet flavon $\phi_{l2}$ in the $\bf{\bar{3}_1}^\prime$ representation, aligned in the same direction as $\phi_l$. In summary, with the invariant terms
\begin{align}
\nonumber
\mathcal{L}_Y &\supset a_e\,\left[\bar{L}^{''}_L \phi_l\right]_{\bf{1}} E^{''3}_{R} + b_e \, \left[ \bar{L}^{''}_L \phi_l\right]_{\bf{2}} E^{''12}_{R} +
c_e \, \left[ \bar{L}^{''}_L \phi_{l2}\right]_{\bf{2}} E^{''12}_{R} \\
&+ a_\nu \, \left[ \bar{L}^{''}_L \phi_\nu \right]_{\bf{2}} \nu^{''12}_{R} +
b_\nu \, \left[ \bar{L}^{''}_L \phi_{\nu2}\right]_{\bf{1}} \nu^{''3}_{R} \,,
\end{align}
the degeneracy of the eigenvalues is lifted as $\left[ \bar{L}_L \phi_l\right]_{\bf{2}} \propto (\bar{L}_{L}^1, \omega_3 \bar{L}_{L}^2)$ whereas $\left[ \bar{L}_{L} \phi_{l2}\right]_{\bf{2}} \propto (\bar{L}_{L}^1, -\omega_3 \bar{L}_{L}^2)$. 
Explicitly, we aim for a normal mass hierachy by picking $(-1,1,0)$ as the additional direction, with a shaping symmetry which distinguishes the neutrino flavons such that each only couples to one of the right-handed neutrino fields. Taking $v_\nu = v_{\nu2}$, the Yukawa term in the neutrino sector then corresponds to
\begin{align}
\nonumber
Y_\nu^{''\dagger} &=  
a_\nu v_{\nu} \left(
\begin{array}{ccc}
\omega_3 & 1 & 0 \\
\omega_3 & 1 & 0 \\
\omega_3 & 1 & 0
\end{array}
\right)
+
b_\nu v_{\nu} \left(
\begin{array}{ccc}
0 & 0 & -1 \\
0 & 0 & 1 \\
0 & 0 & 0
\end{array}
\right)\,,
\end{align}
and therefore we have
\begin{align}
\nonumber
Y_\nu^{''\dagger} Y^{''}_\nu = v_\nu^2
\left(
\begin{array}{ccc}
2 a_\nu^2 + b_\nu^2  & 2 a_\nu^2 - b_\nu^2 & 2 a_\nu^2 \\
2 a_\nu^2 - b_\nu^2 & 2 a_\nu^2 + b_\nu^2 & 2 a_\nu^2 \\
2 a_\nu^2 & 2 a_\nu^2 & 2 a_\nu^2
\end{array}
\right)\,, \,\,\,\,\,\,\,\,\,\,
Y_l^{''\dagger} Y^{''}_l &= v_l^2
\left(
\begin{array}{ccc}
|b_e - c_e|^2 & 0 & 0 \\
0 & |b_e + c_e|^2 & 0 \\
0 & 0 & |a_e|^2
\end{array}
\right)\,,
\end{align}
with $a_\nu$ from the contraction of $(1,1,1)$ with the $\Delta(96)$ doublet and $b_\nu$ from the contraction of $(-1,1,0)$ with the singlet right-handed neutrino, respectively. These map to \eqref{eq:genMassD96} with
\begin{align}
\nonumber
m_{l_2}^2 &\leftrightarrow |b_e - c_e|^2, \,\,\,\,\,\,
m_{l_3}^2 \leftrightarrow |b_e + c_e|^2, \,\,\,\,\,
m_{l_1}^2 \leftrightarrow |a_e|^2, \\
m_{\nu_2}^2 &\leftrightarrow  
|a_\nu|^2,\,\,\,\,\,\,\,\,\,\,\,\,\, \ \ 
m_{\nu_3}^2 \leftrightarrow 
 |b_\nu|^2,\,\,\,\,\,\,\,\,\,\,\,\,\, \ \
m_{\nu_1}^2 = 0,
\end{align}
again up to constant prefactors and VEV, thereby realizing the desired shapes.
\subsubsection*{The Leptoquark Sector}
As seen in Table \ref{tab:NADSresult}, from the bottom-up perspective of the scans in \cite{Bernigaud:2019bfy}, one does not have control over the coupling $x_e$ in \eqref{eq:yukeisolation0} when only lepton symmetries are active.  The $T_l$ symmetry controls the overall shape of the term (electron isolation), but not the quantization of the ratio of $\lambda_{se}/\lambda_{be}$.  This can be seen practically by observing that $\Delta(96)$ is generated by $\Delta(96) \cong \lbrace T^\prime_l, T^\prime_\nu \rbrace$, and neither of these RFS generators knows about $x_e$.  Hence, one derives that in the model basis the generic RFS-invariant leptoquark coupling is given by 
\begin{equation}
\label{eq:D96lqgen}
\lambda_{dl}^{''\dagger}\lambda_{dl}^{''} \overset{!}{=} (1+\vert x_e \vert ^2)\,\vert \lambda_{be}\vert^2 \left(
\begin{array}{ccc}
0 & 0 & 0 \\
0 & 0 & 0 \\
0 & 0 & 1
\end{array}
\right)\,,
\end{equation}
where we observe that, thanks to the Hermitian combination we have constructed, the appearance of $x_e$ in this relationship is not due to the mixing matrix $\Lambda_d$, but instead the mass-eigenstate isolation pattern itself (cf. \eqref{eq:yukeisolation0}).

This term can now be easily built using one of the charged lepton flavons, taking the leptoquark field to transform either as $\bf{1}$ (selects $\phi_l$) or $\bf{1'}$ (selects $\phi_{l2}$). For simplicity we consider the trivial singlet option:
\begin{equation}
\label{eq:D96LQLag}
\mathcal{L} \supset\,\, a_\Delta^i \,\bar{Q}_L^{'' i} [L_L^{''} \phi_{l1}]_{\mathbf{1}} \Delta  \,.
\end{equation}
 As in the $A_4$ models described in \cite{Varzielas:2015iva}, contracting $\left[ L_L \phi_{l1} \right]_1$ gives one of the lepton isolation cases ensuring the leptoquark couples only to one lepton flavour. In this case the VEV in the model building basis is $(0,0,1)$, leading to 
\begin{equation}
\label{eq:D96lqmodel}
\lambda_{dl}^{''} = a_\Delta \left(
\begin{array}{ccc}
0 & 0 & 0 \\
0 & 0 & 0 \\
0 & 0 & 1
\end{array}
\right)\,.
\end{equation}
This is written in an unknown quark basis where the third row corresponds to the specific combination of the three components of $\bar{Q}_L^{''}$ and $a_\Delta$ which is the appropriate function of the three $a_\Delta^i$ coefficients.
In the mass-eigenstate basis, the model yields the electron isolation pattern as expected from the results in our previous paper \cite{Bernigaud:2019bfy}.
To be more precise, we can sum over the uncertainty of the quark sector that we are not controlling with the symmetry. As $\lambda_{dl}^{''}$ only has entries in the third column, the resulting $\lambda_{dl}^{''\dagger}\lambda_{dl}^{''}$ combination only has a non-zero (3,3) entry proportional to the modulus of the third column vector, therefore the model is indeed predicting the structure in \eqref{eq:D96lqgen}.


\subsection{$D_{15}$ Model for $U_{CKM}$, $U_{PMNS}$, and Leptoquarks}
\label{sec:D15}

We now consider a $D_{15}$ model\footnote{Dihedral groups and their double-valued cousins have been favored in the model-building community for some time, see e.g. \cite{Frampton:1994rk} for an early example of the latter.  In fact, the quark sector of the current model we consider can also be mapped to a scan result from \cite{Varzielas:2016zuo}.} that makes predictions for both CKM and PMNS mixing alongside of the ratio of leptoquark couplings denoted by $x_\mu$.  The scan result from \cite{Bernigaud:2019bfy} is repeated in Table \ref{tab:NADSresult}, whose first column reveals that a hexagonal PMNS matrix $U_{HM}$ is predicted alongside of Cabibbo mixing with $\theta_C = \pi/15$ for the CKM matrix, while the second through fourth columns reveal the following symmetry-breaking pattern:
\begin{equation}
\label{eq:D15diag}
\mathcal{G_{F}} \cong D_{15}   \rightarrow	  \mathcal{G_{\nu,\text{l},\text{u},\text{d}}} \cong \mathbb{Z}_2 \,.
\end{equation}
From Table \ref{tab:NADSresult} we can immediately construct the leptoflavour-basis RFS generators with \eqref{Eq:LGenerators_Flavour_basis}, finding that the neutrino and up-quark matrix have non-trivial structure in all three matrix sectors.

As above, we attempt to find a basis within which its easy to manipulate the relevant $D_{15}$ group product rules.  To that end, we consider the following unitary transformation that block diagonalizes the leptoflavour-basis generators:
\begin{equation} 
\label{eq:D96pmatrix}
P = \begin{pmatrix}
\frac{1}{\sqrt{2}} & \frac{1}{\sqrt{2}} & 0 \\
\frac{1}{\sqrt{2}} & -\frac{1}{\sqrt{2}} & 0 \\
0 & 0 & 1
\end{pmatrix}
\begin{pmatrix}
1 & 0 & 0 \\
0 & 0 & 1 \\
0 & 1 & 0
\end{pmatrix} =
\begin{pmatrix}
\frac{1}{\sqrt{2}} & 0 & \frac{1}{\sqrt{2}} \\
\frac{1}{\sqrt{2}} & 0 & -\frac{1}{\sqrt{2}} \\
0 & 1 & 0
\end{pmatrix}\,.
\end{equation}
Applying $P$ with $T_{f}^{\prime \prime} = P^\dagger T_{f}^\prime P$ we get the following expressions for the RFS-generators in the model basis:
\begin{align}
\begin{split}
 T_\nu^{\prime \prime}
 = \begin{pmatrix}
-1 & 0 & 0 \\
0 & \frac{1}{2} & \frac{\sqrt{3}}{2}\\
0 & \frac{\sqrt{3}}{2} & -\frac{1}{2}
\end{pmatrix}, \,\,\,\,\,
T_{u}^{\prime \prime}  
= \begin{pmatrix}
-1 & 0 & 0 \\
0 &\cos \frac{2\pi}{15} & \sin \frac{2\pi}{15}\\
0 &\sin \frac{2\pi}{15} & -\cos \frac{2\pi}{15}
\end{pmatrix}, \,\,\,\,\,
T_{l,d}^{\prime \prime}  
= \begin{pmatrix}
-1 & 0 & 0\\
0 & 1 & 0\\
0 & 0 & -1
\end{pmatrix}\,.
\end{split}
\label{Left_RFS_in_Model}
\end{align}
We now want to identify these matrices with $D_{15}$ generating elements in certain irreducible representations, and multiple sources are available that catalogue properties of the dihedral series $D_{N}$.  We have distilled the relevant information specific to $D_{15}$ in Appendix \ref{app:D15}, from which we see that $T_{u,d,l,\nu}^{\prime \prime}$ can be easily expressed in terms of the group elements ${\bf{a}}$ and ${\bf{b}}$ for the combination $\mathbf{1_- + 2_1}$ and $\mathbf{1_- + 2_5}$:
\begin{equation}
\begin{split}
&T_{d}^{\prime \prime} = \mathbf{b}, \ \ \ T_{u}^{\prime \prime} = \mathbf{ab} \ \ \ \ \ \  \text{for} \ \ \ \ \ \mathbf{1_- + 2_1}\,, \\
&T_{l}^{\prime \prime} = \mathbf{b}, \ \ \ T_{\nu}^{\prime \prime} = \mathbf{ab} \ \ \ \ \ \  \text{for} \ \ \ \ \ \mathbf{1_- + 2_5}\,.
\end{split}
\label{Preserved_elements}
\end{equation}
Here $\bf{1_{-}}$ is the lone non-trivial $D_{15}$ singlet, and $\bf{2_{1,5}}$ denote two of the seven doublets of the group.  Critically, we observe that \eqref{Left_RFS_in_Model}-\eqref{Preserved_elements} indicate that $\bf{2_1}$ ($\bf{2_5}$) is the appropriate $D_{15}$ charge for the two flavons $\phi_{u,d}$ ($\phi_{l,\nu}$) that we introduce according to the algorithm in Section \ref{sec:THEORY}, and we can use \eqref{Left_RFS_in_Model} to work out the expressions for these doublet VEV, finding
\begin{equation}
\langle \phi_\nu \rangle = v_\nu \left( 1, \sqrt{3} \right)^T, \ \ \ \langle \phi_u \rangle = v_u \left(  \cos \frac{\pi}{15}, \sin \frac{\pi}{15} \right)^T, \ \ \ \langle \phi_{d,l} \rangle = v_{d,l} \left( 1, 0 \right)^T\,.
\label{vevs}
\end{equation}
Finally, one derives that in the absence of mixing ambiguities, the model-basis mass matrices are given by
\begin{align}
\nonumber
m_{\nu}^{''}&\overset{!}{=}
\left( 
\begin{array}{ccc}
m_{\nu_3} & 0 & 0 \\
0 & \frac{1}{4}\left(3 m_{\nu_1} + m_{\nu_2} \right) & \frac{ \sqrt{3}}{4} \left(m_{\nu_2} - m_{\nu_1} \right)  \\
0 &  \frac{ \sqrt{3}}{4} \left(m_{\nu_2} - m_{\nu_1} \right)  &  \frac{1}{4}\left(m_{\nu_1} + 3 m_{\nu_2} \right)
\end{array}
\right), \\
\nonumber
m_{l}^{''\dagger}m_{l}^{''}&\overset{!}{=}
\left( 
\begin{array}{ccc}
\frac{1}{2} \left( m_{l_2}^2 + m_{l_3}^2\right) & 0 & \frac{1}{2} \left(m_{l_2}-m_{l_3}\right)\left(m_{l_2}+m_{l_3}\right) \\
0 & m_{l_1}^2 & 0 \\
\frac{1}{2} \left(m_{l_2}-m_{l_3}\right)\left(m_{l_2}+m_{l_3}\right) & 0 & \frac{1}{2} \left( m_{l_2}^2 + m_{l_3}^2\right) 
\end{array}
\right), \\
\nonumber
m_{u}^{''\dagger}m_{u}^{''}&\overset{!}{=}
\left( 
\begin{array}{ccc}
m_{u_3}^2 & 0 & 0 \\
0 & m_{u_1}^2 \cos^2\frac{\pi}{15} + m_{u_2}^2 \sin^2\frac{\pi}{15} & \frac{1}{2}\left(m_{u_1}-m_{u_2} \right) \left(m_{u_1}+m_{u_2} \right) \sin \frac{2\pi}{15} \\
0 & \frac{1}{2}\left(m_{u_1}-m_{u_2} \right) \left(m_{u_1}+m_{u_2} \right) \sin \frac{2\pi}{15} & m_{u_2}^2 \cos^2\frac{\pi}{15} + m_{u_1}^2 \sin^2\frac{\pi}{15}
\end{array}
\right), \\
\label{eq:D15genMass}
m_{d}^{''\dagger}m_{d}^{''}&\overset{!}{=}
\left( 
\begin{array}{ccc}
m_{d_3}^2 & 0 & 0 \\
0 & m_{d_1}^2 & 0 \\
0 & 0 & m_{d_2}^2
\end{array}
\right),
\end{align}
where $m_{A_i}$ are the associated mass eigenvalues, and where we have used dagger combinations for the charged fermions to remove the dependence on RH transformations.  However, unlike the $\Delta(96)$ model of Section \ref{sec:D96}, we see from Table \ref{tab:NADSresult} that all of the RFS generators have degenerate eigenvalues, and hence there are again freedoms in the associated mass and mixing matrices thanks to \eqref{eq:degenEV}.  We will discuss these when they become relevant below.

\subsubsection*{The Quark Sector}
The results in \eqref{Left_RFS_in_Model}-\eqref{eq:D15genMass} strongly indicate that 
the second and third generations of LH quarks should transform as a $D_{15}$ doublet, while the first generation of up quarks transforms as a non-trivial singlet.  Similarly, the first and third generations of RH up and down quarks should transform as a non-trivial singlet, while the second generation of both families transforms trivially. Furthermore, \eqref{Preserved_elements} indicates that the flavons $\phi_{d,u}$ associated to these sectors should transform as a ${\bf{2_1}}$ under $D_{15}$, a fact that helped us derive \eqref{vevs}.  This information is summarized in Table \ref{Tab:D15AQuantumNumbers}.  

\begin{table}[t]
    \centering
    \renewcommand{\arraystretch}{1.5}
    \begin{tabular}{|c||c|c|c|c|c|c|c|c|c|c|c|}
    \hline
  \  & $Q^{''1}$ & $Q^{''23}$ & $u_{R}^{''1}$ & $u_{R}^{''2}$ & $u_{R}^{''3}$ & $d_{R}^{''1}$ & $d_{R}^{''2}$ & $d_{R}^{''3}$  & $\phi_u$ & $\phi_d$ \\
  \hline
  \hline
  $D_{15}$ & $\mathbf{1_-}$ & $\mathbf{2}_{\bf{1}}$ & $\mathbf{1_-}$ & $\mathbf{1}$ & $\mathbf{1_-}$ & $\mathbf{1_-}$ & $\mathbf{1}$ & $\mathbf{1_-}$ & $\mathbf{2_1}$ & $\mathbf{2_1}$ \\
  \hline
\end{tabular}    
\caption{Representations of the quarks under $D_{15}$}
    \label{Tab:D15AQuantumNumbers}
\end{table}

With an appropriate shaping symmetry preventing $\phi_{d,u}$ from coupling to undesirable sectors as well as distinguishing $u_{R}^{''1}$ from $u_{R}^{''3}$ and $d_{R}^{''1}$ from $d_{R}^{''3}$,\footnote{However, we will soon see that off-diagonal entries in $Y_d^{''\dagger}$ will become desirable once we begin to discuss the leptoquark sector below, and therefore the implied shaping symmetry present in \eqref{eq:D15quarkLag} will be modified to allow such additional operators.} one can quickly obtain the model-basis Yukawa sector for the quarks using Table \ref{Tab:D15AQuantumNumbers},
\begin{equation}
\begin{split}
\mathcal{L}_{Y} &\supset a_u \, \bar{Q}^{''1}_L u_{R}^{''1} + b_u \,\left[\bar{Q}^{''23}_L \phi_u\right]_{\mathbf{1}}u_{R}^{''2} + c_u\, \left[\bar{Q}^{''23}_L \phi_u\right]_{\mathbf{1_-}}u_{R}^{''3} \\
\label{eq:D15quarkLag}
&+  a_d \, \bar{Q}^{''1}_L d_{R}^{''1} + b_d \, \left[\bar{Q}^{''23}_L \phi_d\right]_{\mathbf{1}}d_{R}^{''2} + c_d \, \left[\bar{Q}^{''23}_L \phi_d\right]_{\mathbf{1_-}}d_{R}^{''3}\,,
\end{split}
\end{equation}
where Higgs fields and scale suppressions are again ommitted.  Using the VEV from \eqref{vevs} and product rules from Appendix \ref{app:D15}, we get the following Yukawa matrices 
\begin{equation}
Y^{''\dagger}_u = v_u\begin{pmatrix}
a_u/v_u & 0 & 0 \\
0 & b_u \cos \frac{\pi}{15} & -c_u \sin \frac{\pi}{15}  \\
 0 & b_u \sin \frac{\pi}{15} & c_u \cos \frac{\pi}{15}  \\

\end{pmatrix}, \ \ \ \ \ 
Y^{''\dagger}_d = v_d \begin{pmatrix}
a_d/v_d & 0 & 0 \\
0 & b_d & 0 \\
0 & 0 & c_d
\end{pmatrix}.
\end{equation}
Assembling these into their Hermitian combinations, one arrives at
\begin{align}
\nonumber
Y_u^{\prime \prime \dagger} Y_u^{\prime \prime} &= v_u^2 \left(
\begin{array}{ccc}
\vert a_u \vert^2/v_u^2 & 0 & 0 \\
0 & \vert b_u \vert^2 \cos^2 \frac{\pi}{15} +  \vert c_u \vert^2 \sin^2 \frac{\pi}{15} & \frac{1}{2} \left(\vert b_u \vert^2 - \vert c_u \vert^2 \right)\sin \frac{2\pi}{15} \\
0 & \frac{1}{2} \left(\vert b_u \vert^2 - \vert c_u \vert^2 \right)\sin \frac{2\pi}{15} & \vert c_u \vert^2 \cos^2 \frac{\pi}{15} +  \vert b_u \vert^2 \sin^2 \frac{\pi}{15} 
\end{array}
\right)\,, \\
\label{eq:D15Yd}
Y_d^{\prime \prime \dagger}Y_d^{\prime \prime} &= v_d^2 \left(
\begin{array}{ccc}
\vert a_d \vert^2/v_d^2 & 0 & 0 \\
0 & \vert b_d \vert^2 & 0 \\
0 & 0 & \vert c_d \vert^2
\end{array}
\right)\,,
\end{align}
which maps, up to prefactors and VEV, to \eqref{eq:D15genMass} with
\begin{align}
\nonumber
\,\vert a_u \vert^2 \leftrightarrow \vert m_{u_3} \vert^2, \,\,\,\,\,\vert b_u \vert^2 \leftrightarrow \vert m_{u_1} \vert^2,  \,\,\,\,\, \vert c_u \vert^2 \leftrightarrow \vert m_{u_2} \vert^2  ,
\end{align}
and analogous relations for the mapping of $Y_d$.

\subsubsection*{The Lepton Sector}
Similarly, the matrices in \eqref{Left_RFS_in_Model}-\eqref{eq:D15genMass} suggest that the second and third LH generations of SU(2) doublet leptons transform as a ${\bf{2_5}}$ $D_{15}$ doublet, along with the associated flavons $\phi_{\nu,l}$. The $L_L^{''1}$ and first and third generations of $E_R^{''}$ are to be charged as non-trivial singlets, while $E_R^{''2}$ transforms trivially.
\begin{table}[t]
    \centering
    \renewcommand{\arraystretch}{1.5}
    \begin{tabular}{|c||c|c|c|c|c|c|c|c|c|c|c|}
    \hline
  \  & $L^{''1}$ & $L^{''23}$ & $E_{R}^{''1}$ & $E_{R}^{''2}$ & $E_{R}^{''3}$ & $\phi_\nu$ & $\phi_l$ \\
  \hline
  \hline
  $D_{15}$ & $\mathbf{1_-}$ & $\mathbf{2_5}$ & $\mathbf{1_-}$ & $\mathbf{1}$ & $\mathbf{1_-}$ & $\mathbf{2_5}$ & $\mathbf{2_5}$ \\
  \hline
\end{tabular}    
\caption{Representations of the leptons under $D_{15}$}
    \label{Tab:D15AQuantumNumbers_Leptons}
\end{table}
Using this information, assembled in Table \ref{Tab:D15AQuantumNumbers_Leptons}, one reconstructs the LO Lagrangian as
\begin{equation}
\begin{split}
\mathcal{L}_{Y} &\supset a_e \, \bar{L}^{''1}_L E_{R}^{''1} + b_e \, \left[ \bar{L}^{''23}_L \phi_l \right]_{\mathbf{1}}E_{R}^{''2} + c_e \, \left[ \bar{L}^{''23}_L \phi_l \right]_{\mathbf{1_-}}E_{R}^{''3} + d_e \, \bar{L}^{''1}_L E_{R}^3 + \epsilon_e \, \left[ \bar{L}^{''23}_L \phi_l \right]_{\mathbf{1_-}}E_{R}^{''1}\\
&+ a_\nu \bar{L}^{c\,''1}_L L_L^{''1} +  b_\nu \left[ \bar{L}^{c\,''23}_L \phi_\nu\right]_\mathbf{1}\left[ \bar{L}^{''23}_L \phi_\nu\right]_\mathbf{1} + c_\nu \left[ \bar{L}^{c\,''23}_L \phi_\nu\right]_\mathbf{1_-}\left[ \bar{L}^{''23}_L \phi_\nu\right]_\mathbf{1_-} \,.
\end{split}
\end{equation}
We quickly derive the following terms for neutrino masses and charged lepton Yukawas
\begin{equation}
m^{''}_\nu = v_\nu^2 \begin{pmatrix}
a_\nu/v_\nu^2 & 0 & 0 \\
0 & b_\nu + 3c_\nu & \sqrt{3}(b_\nu-c_\nu) \\
0 & \sqrt{3}(b_\nu-c_\nu) & 3b_\nu + c_\nu
\end{pmatrix}, \ \ \ \ 
Y^{''\dagger}_l = v_l \begin{pmatrix}
a_e/v_l & 0 & d_e/v_l\\
0 & b_e & 0 \\
\epsilon_e & 0 & c_e
\end{pmatrix},
\end{equation}
which gives
\begin{equation}
\label{eq:D15modelYl2b}
Y_l^{\prime \prime \dagger} Y_l^{\prime\prime} = v_l^2 \left(
\begin{array}{ccc}
(\vert a_e \vert ^2 + \vert d_e \vert ^2)/v_l^2 & 0 & (d_e c_e^\star + \epsilon_e^\star a_e)/v_l \\
0 & \vert b_e \vert ^2 & 0 \\
(\epsilon_e a_e^\star + d_e^\star c_e)/v_l & 0 & \vert c_e \vert^2 + \vert \epsilon_e \vert ^2 
\end{array}
\right)\,.
\end{equation}
While the neutrino mass maps directly to \eqref{eq:D15genMass}, the charged lepton term apparently does not without additional fine tuning of the parameters.  For example, setting $d_e = \epsilon_e$ and $c_e = a_e$, one can recover the corresponding term in \eqref{eq:D15genMass}, and only then will a diagonal matrix of eigenvalues be returned for $Y_l^{\dagger}Y_l$ upon (un)rotating \eqref{eq:D15modelYl2b} to the mass-eigenstate basis with $\Lambda_l$ and $P$.  However, we simultaneously observe that the coupling $Y_l^{\prime \prime \dagger} Y_l^{\prime\prime}$ still respects the required RFS invariance,
\begin{equation}
\label{eq:RFSsurprise}
T_l^{\prime\prime\dagger}\,Y_l^{\prime \prime \dagger} Y_l^{\prime\prime}\,T_l^{\prime\prime} = Y_l^{\prime \prime \dagger} Y_l^{\prime\prime}\,.
\end{equation}
Are these claims contradictory?  After all, we argued that successfully mapping to \eqref{eq:D15genMass} is a sufficient condition for ensuring that the EFT yields the desired IR phenomenology and RFS symmetry-breaking patterns, and while it appears that \eqref{eq:D15modelYl2b} cannot do so without unappealing assumptions, the RFS invariance still holds in \eqref{eq:RFSsurprise}.  

The solution to this puzzle resides in the fact that, as in the A$_4$ Altarelli-Feruglio case, \eqref{eq:D15genMass} does \emph{not}, in fact, represent the most generic set of RFS-invariant mass matrices, and we will now show that the apparent fine-tuning required in mapping \eqref{eq:D15modelYl2b} to \eqref{eq:D15genMass} can be understood as a top-down manifestation of \eqref{eq:degenEV}.

We begin by recalling that the $T_l^{\prime\prime}$ generator cannot distinguish between $\Lambda_l$ and $\Lambda_l \cdot R_{23}$,
\begin{equation}
T_l^{\prime \prime} = \tilde{T}_{l} = P^\dagger \Lambda_l R_{23} T_l R_{23}^\dagger\Lambda_l^\dagger P,
\end{equation}
where our tilde notation indicates that this can be understood a basis change on the charged lepton field, such that (starting from the mass-eigenstate basis), we are now operating with
\begin{equation}
\label{eq:lLchange}
l_L \rightarrow R_{23}^\dagger \, \Lambda_{l}^{\dagger} \,P \,\tilde{l}_L\,,
\end{equation}
and so the most generic mass matrix invariant under the $\tilde{T}_{l} = T_l^{\prime \prime}$ generator that our algorithm to find $D_{15}$ knows about is given by
\begin{equation}
\tilde{m}^\dagger_l \tilde{m}_l = P^\dagger \Lambda_L R_{23} \, m_l^\dagger m_l \, R_{23}^\dagger \Lambda_l^\dagger P \,\,\, \Longleftrightarrow \,\,\, \tilde{T}_{l}^\dagger \tilde{m}^\dagger_l \tilde{m}_l \tilde{T}_l = \tilde{m}^\dagger_l \tilde{m}_l =T_{l}^{\prime\prime \dagger} \tilde{m}^\dagger_l \tilde{m}_l T^{\prime\prime}_l 
\end{equation}
with the the mass matrix elements given by 
\begin{align}
\nonumber
\left(\tilde{m}^\dagger_l \tilde{m}_l \right)^{11} &= \frac{1}{2}\left(m_{l_2}^2 + m_{l_3}^2 + \left(m_{l_3}^2 - m_{l_2}^2\right) \cos \delta_{23} \sin 2\theta_{23} \right), \\
\nonumber
\left(\tilde{m}^\dagger_l \tilde{m}_l \right)^{13} &=\frac{1}{2}\left(m_{l_2}^2 - m_{l_3}^2\right)  \left(\cos 2\theta_{23} - i \sin 2 \theta_{23} \sin \delta_{23} \right), \\
\nonumber
\left(\tilde{m}^\dagger_l \tilde{m}_l \right)^{22} &=  m_{l_1}^2,\\
\nonumber
\left(\tilde{m}^\dagger_l \tilde{m}_l  \right)^{31} &= \frac{1}{2} \left(m_{l_2}^2 - m_{l_3}^2\right) \left(\cos 2\theta_{23} + i \sin 2 \theta_{23} \sin \delta_{23} \right),\\
\nonumber
\left(\tilde{m}^\dagger_l \tilde{m}_l \right)^{33} &= \frac{1}{2}\left(m_{l_2}^2 + m_{l_3}^2 + \left(m_{l_2}^2 - m_{l_3}^2\right)  \cos \delta_{23} \sin 2\theta_{23} \right), \\
\label{eq:RFSgenMD30}
\left(\tilde{m}^\dagger_l \tilde{m}_l  \right)^{12} &= \left(\tilde{m}^\dagger_l \tilde{m}_l \right)^{21} = \left(\tilde{m}^\dagger_l \tilde{m}_l  \right)^{23} = \left(\tilde{m}^\dagger_l \tilde{m}_l  \right)^{32} = 0,
\end{align}
and where $\lbrace \delta, \theta \rbrace_{23}$ denote the free parameters our formalism has no control over.

However, \eqref{eq:lLchange} also implies that a diagonal charged current in this basis (we have not applied a change on any other field) implies a modified PMNS matrix in the (physical) mass-eigenstate basis:
\begin{equation}
\overline{\tilde{l}}_L \gamma_\mu \tilde{\nu}_L W^-_\mu \,\,\,\,\, \underset{\text{mass basis}}{\longleftrightarrow}\,\,\,\,\, \overline{l}_L \gamma_\mu \underbrace{R_{23}^\dagger \, U_{HM}}_{U_{PMNS}} \, \nu_L W^{-}_\mu\,.
\end{equation}
This is consistent with the well-known fact that, in a more generic flavour basis, a degeneracy in $T_l$ should translate to a free parameter in $U_l$ and therefore also $U_{PMNS}$.

Now, one might be tempted to conclude that we should also translate $\nu_L$ with a compensating factor of $R_{23}$, so that the definition of the PMNS is preserved a la
\begin{equation}
\label{eq:nuChange}
\nu_L \rightarrow U_{HM}^\dagger R_{23}^\dagger \Lambda_l^\dagger P \tilde{\nu}_L \,\,\, \Longrightarrow \,\,\,\overline{\tilde{l}}_L \gamma_\mu \tilde{\nu}_L W^-_\mu \,\,\,\,\, \underset{\text{mass basis}}{\longleftrightarrow}\,\,\,\,\, \overline{l}_L \gamma_\mu \, \underbrace{U_{HM}}_{U_{PMNS}} \, \nu_L W^{-}_\mu \, .
\end{equation}
However, this corresponds to a generic neutrino mass matrix given by 
\begin{equation}
\tilde{m}_\nu = P^T\Lambda_{l}^* R_{23}^\star U_{HM}^* m_\nu U_{HM}^\dagger R_{23}^\dagger \Lambda_{l}^\dagger P\,,
\end{equation}
which is left invariant under 
\begin{equation}
\tilde{T}_\nu = P^\dagger \Lambda_l R_{23} U_{HM} T_{\nu} U_{HM}^\dagger R_{23}^\dagger \Lambda_l^\dagger P\neq T_\nu^{\prime \prime} \,.
\end{equation}
That is, the neutrino generator \emph{knows} about this basis change, which differs from our original observation that $D_{15}$ doesn't know the difference between $T_l^{\prime \prime}$ and $\tilde{T}_{l}$.  Indeed, we have checked that (at least at certain values of $\delta_{23}$ and $\theta_{23}$) the group generated by $ \mathcal{G_F} \cong \lbrace \tilde{T}_{l}, \tilde{T}_\nu, \tilde{T}_u, \tilde{T}_d \rbrace$ is \emph{not} $D_{15}$, and may not even be finite!\footnote{Note that our tilde notation here does not imply that we have performed a similarity transformation on all four elements of the $\prime\prime$ generating set, but rather that upon making the basis transformations in \eqref{eq:lLchange}-\eqref{eq:nuChange} we then derive a set of RFS generators from the Lagrangian where $\tilde{T}_{u,d,l} = T^{\prime \prime}_{u,d,l}$ but $\tilde{T}_{\nu} \neq T^{\prime \prime}_{\nu}$.  Hence the tilde set can easily generate a different group, as we have seen.}  So making this compensating change in \eqref{eq:nuChange} is inconsistent with the starting point of our analysis, and cannot be done. 

In conclusion, moving to the tilde basis requires no more work from the model building side, but implies a generalized RFS-invariant charged lepton mass matrix, and therefore a generalized prediction for the physical PMNS matrix given by $U_{PMNS} = R_{23}^\dagger \cdot U_{HM}$.  The parameters of $R_{23}$ are therefore functions of the (unspecified) coupling strengths of the EFT operators, since when solving the system of equations implied by mapping \eqref{eq:RFSgenMD30} to \eqref{eq:D15modelYl2b}, one easily sees that 
\begin{align}
\nonumber
m_{l_1}^2 &\leftrightarrow \vert b_e \vert^2\,, \\
\nonumber
m_{l_2}^2 &\leftrightarrow \left(\vert d_e \vert^2 + \vert \epsilon_e \vert^2 + \vert a_e \vert^2 + \vert c_e \vert^2 +\left(\vert \epsilon_e \vert^2 + \vert c_e \vert^2 - \vert a_e \vert^2 - \vert d_e \vert^2\right)\csc 2 \theta_{23} \sec \delta_{23}\right) \,,\\
\label{eq:R23massesD15}
m_{l_3}^2 &\leftrightarrow \left(\vert d_e \vert^2 + \vert \epsilon_e \vert^2 + \vert a_e\vert^2 + \vert c_e \vert^2 +\left(\vert d_e \vert^2 + \vert a_e \vert^2 - \vert c_e \vert^2 - \vert \epsilon_e \vert^2\right)\csc 2 \theta_{23} \sec \delta_{23}\right) \,,
\end{align}
up to VEV and prefactors, with
\begin{align}
\nonumber
\delta_{23} &= \arctan \left(\mathrm{Im}\left[  \frac{2(d_e c_e^* + \epsilon_e^* a_e)}{|a_e|^2 + |d_e|^2 - |c_e|^2 - |\epsilon_e|^2}  \right]\right)\,,\\
\label{eq:R23D15}
2\theta_{23} &= \arctan \left(-\mathrm{Re}\left[  \frac{2(d_e c_e^* + \epsilon_e^* a_e)\cos\delta_{23}}{|a_e|^2 + |d_e|^2 - |c_e|^2 - |\epsilon_e|^2}  \right]^{-1}\right)\,.
\end{align}
Hence, while no fine-tuning is required in achieving this map,  the model's prediction for $U_{PMNS}$ is ambiguous up to the quantization of the couplings $\lbrace a_e, ..., \epsilon_e\rbrace$, which may result from a higher UV symmetry (e.g. a GUT) that relates the otherwise independent operators of the EFT.  Such an attempt is obviously well beyond our scope in this paper.  However, thinking from a more phenomenological perspective, one can instead fit the parameters $\lbrace \theta_{23}, \delta_{23}\rbrace$ to available experimental data for the PMNS matrix, which then implies relationships amongst the top-down model's parameters, according to \eqref{eq:R23massesD15}-\eqref{eq:R23D15}.  Regardless, we see clearly that the ignorance of $R_{23}$ in the \emph{bottom-up} RFS generation of $\mathcal{G_F}$ has consistently manifested itself in a certain lack of predictivity in the \emph{top-down} EFT.

\subsubsection*{The Leptoquark Sector}
The naive RFS-invariant leptoquark coupling for the $d-l$ operator expected in the model basis is given by 
\begin{equation}
\label{eq:lamdlNFP}
\lambda_{dl}^{\prime \prime} \overset{!}{=} \frac{\lambda_{b\mu}}{\sqrt{2}} \left(
\begin{array}{ccc}
1 & 0 & 1 \\
0 & 0 & 0 \\
1 & 0 & 1
\end{array}
\right)\,,
\end{equation}
This can be achieved using the following Lagrangian
\begin{equation}
\label{eq:D15LQLag}
\mathcal{L} \supset\,\, a_\Delta \,\bar{Q}_L^{''1} L_L^{''1} + b_\Delta \, \bar{Q}^{''1}_L \left[L_L^{''23} \phi_l\right]_{\mathbf{1_-}} + c_\Delta \, \left[\bar{Q}^{''23}_L \phi_d\right]_{\mathbf{1_-}} \left[L_L^{''23} \phi_l\right]_{\mathbf{1_-}} + d_\Delta \, \left[\bar{Q}^{''23}_L \phi_d\right]_{\mathbf{1_-}} L_L^{''1} \,,
\end{equation}
but again only with an additional tuning of the parameters,
\begin{equation}
a_\Delta  = b_\Delta  = c_\Delta = d_\Delta  \ \  \Longrightarrow \ \ \lambda^{\prime \prime}_{dl} = \begin{pmatrix}
a_\Delta & 0 & a_\Delta \\
0  & 0 & 0 \\
a_\Delta & 0 & a_\Delta
\end{pmatrix}.
\end{equation}
However, following the above discussion for charged leptons, \eqref{eq:lamdlNFP} is modified when we consider the $R_{23}$ free rotation yielding \eqref{eq:RFSgenMD30}, and results in the matrix 
\begin{equation}
\label{eq:lamdlFPa}
\tilde{\lambda}_{dl} \overset{!}{=} \frac{\lambda_{b\mu}}{\sqrt{2}} \left(
\begin{array}{ccc}
\mathcal{E}^l_{--} & 0 & \mathcal{E}^l_{+-}  \\
0 & 0 & 0 \\
\mathcal{E}^l_{--}  & 0 & \mathcal{E}^l_{+-} 
\end{array}
\right)\,, \,\,\,\, \mathcal{E}^{l}_{\pm \pm} \equiv \left(\cos \theta^l_{23}\, \pm \, e^{\pm\, i \,\delta^l_{23}} \sin \theta^l_{23}\right) \,,
\end{equation}
whose first and third columns, corresponding to lepton generations, are now distinguished.\footnote{ Note that in the definition of $\mathcal{E}^{l}_{\pm \pm}$, the two $\pm$ labels correspond to the first and second $\pm$ appearing on the RHS, respectively, while the $l,d$ superscript denotes parameters from the lepton and down quark sectors.} Hence the need to fine-tune parameters between them disappears, although the otherwise independent couplings $\lbrace a_\Delta, b_{\Delta}, c_{\Delta}, d_{\Delta} \rbrace$ and $\lbrace a_e, b_e, c_e, d_e \rbrace$ are linked through \eqref{eq:R23D15} --- they need to be simultaneously fit to functions of the same physical PMNS parameters, and hence are quite correlated. 

Continuing, we also now note that the symmetry structure exposed in \cite{Bernigaud:2019bfy} (cf. Table \ref{tab:NADSresult}) also permits a rotation in the (2,3) sector of $T_d$, and the conversation above as regards the corresponding basis change on the charged-lepton field can be had equally for the down quark, resulting in a further modified \eqref{eq:lamdlFPa} which also distinguishes rows (quark generations),
\begin{equation}
\label{eq:lamdlFPb}
\tilde{\tilde{\lambda}}_{dl}  \overset{!}{=}
\frac{\lambda_{b\mu}}{\sqrt{2}} \left(
\begin{array}{ccc}
\mathcal{E}^d_{--}\,\mathcal{E}^l_{--} & 0 & \mathcal{E}^d_{--}\,\mathcal{E}^l_{+-}  \\
0 & 0 & 0 \\
\mathcal{E}^d_{++}\,\mathcal{E}^l_{--}  & 0 & \mathcal{E}^d_{++}\, \mathcal{E}^l_{+-} 
\end{array}
\right)\,, 
\end{equation}
with $\mathcal{E}^d_{\pm\pm}$ defined analogously to $\mathcal{E}^l_{\pm\pm}$. Then no fine-tuning of the model's couplings in \eqref{eq:D15LQLag} will be required since the mapping
\begin{equation}
a_\Delta \leftrightarrow \lambda_{b\mu} \,\mathcal{E}^d_{--}\,\mathcal{E}^l_{--} , \,\,\,\,\, b_\Delta \leftrightarrow \lambda_{b\mu}\, \mathcal{E}^d_{--}\,\mathcal{E}^l_{+-}, \,\,\,\,\, c_{\Delta} \leftrightarrow  \lambda_{b\mu}\,  \mathcal{E}^d_{++}\,\mathcal{E}^l_{--} , \,\,\,\,\, d_{\Delta} \leftrightarrow  \lambda_{b\mu}\, \mathcal{E}^d_{++}\, \mathcal{E}^l_{+-}   
\end{equation}
is achieved, although again at the expense of the phenomenology not being uniquely nailed down.  But we must also then account for the fact that the $R_{23}$ rotation in the quark sector simulatneously changes the prediction for the associated RFS-invariant mass matrix to
\begin{equation}
\label{eq:D15alteredDmass}
\tilde{\tilde{m}}_d^\dagger\tilde{\tilde{m}}_d \overset{!}{=} \left(
\begin{array}{ccc}
m_{d_3}^2 \cos^2 \theta^q_{23} + m_{d_2}^2 \sin^2 \theta_{23}^q & 0 & -e^{i \delta^q_{23}} \left(m_{d_2}^2 - m_{d_3}^2 \right) \cos \theta_{23}^q \sin \theta_{23}^q \\
0 & m_{d_1}^2 & 0 \\
e^{-i \delta^q_{23}} \left(m_{d_3}^2 - m_{d_2}^2 \right) \cos \theta_{23}^q \sin \theta_{23}^q & 0 & m_{d_2}^2 \cos^2 \theta_{23}^q + m_{d_3}^2 \sin^2 \theta_{23}^q
\end{array}
\right)\,,
\end{equation}
which further corresponds to an altered CKM prediction,
\begin{equation}
\label{eq:D15alteredCKM}
U_{CKM} \rightarrow \left( 
\begin{array}{ccc}
\cos \frac{\pi}{15} & \cos \theta^q_{23} \sin \frac{\pi}{15} & e^{-i \delta^q_{23}} \sin \theta^q_{23} \sin \frac{\pi}{15} \\
-\sin \frac{\pi}{15} &  \cos \theta^q_{23} \cos \frac{\pi}{15}& e^{-i \delta^q_{23}} \sin \theta^q_{23} \cos \frac{\pi}{15} \\
0 & - e^{i \delta_{23}^q} \sin \theta^q_{23} & \cos \theta^q_{23}
\end{array}
\right)\,.
\end{equation}
These changes appear problematic at first sight, since we have already successfully achieved the desired mapping from \eqref{eq:D15Yd} to \eqref{eq:D15genMass}, i.e. the CKM prediction with no mixing ambiguity.  Indeed, the Lagrangian in \eqref{eq:D15quarkLag} does not have operators that can map to the additional contributions in the $(1,3)$ sector of \eqref{eq:D15alteredDmass}. However, we recall that the implicit assumption in building \eqref{eq:D15quarkLag} (and all other effective $\mathcal{L}_Y$ in this paper) is that unspecified shaping symmetries forbid undesirable operators from contributing to the Yukawa.  So, once \eqref{eq:D15alteredDmass} is the appropriate term to be recovered in the Lagrangian of \eqref{eq:D15quarkLag}, one can then assume a different shaping symmetry, such that further operators contribute in a way that allows for a one-to-one mapping.  For example, one can easily obtain the modified quark Lagrangian as the one in \eqref{eq:D15quarkLag} with additional down quark terms which are in fact invariant under $D_{15}$ (given that $d_{R}^{''1}$ and $d_{R}^{''3}$ are not in fact distinguished by $D_{15}$). The down sector would then be
\begin{equation}
\mathcal{L}_{Y} \supset a_d \, \bar{Q}^{''1}_L d_{R}^{''1} + b_d \, \left[\bar{Q}^{''23}_L \phi_d\right]_{\mathbf{1}}d_{R}^{''2} + c_d \, \left[\bar{Q}^{''23}_L \phi_d\right]_{\mathbf{1_-}}d_{R}^{''3}
+
a'_d \, \bar{Q}^{''1}_L d_{R}^{''3}
+
c'_d \, \left[\bar{Q}^{''23}_L \phi_d\right]_{\mathbf{1_-}}d_{R}^{''1}\,,
\end{equation}
where the $a'_d$ and $c'_d$ terms are clearly the terms with the unprimed couplings after undergoing a swap of $d_{R}^{''1}$ and $d_{R}^{''3}$. It is simple to see they create entries in the mass matrix that will allow a successful map to \eqref{eq:D15alteredDmass}:
\begin{equation}
\tilde{\tilde{Y}}^{\dagger}_d = v_d \begin{pmatrix}
a_d/v_d & 0 & a'_d/v_d \\
0 & b_d & 0 \\
c'_d & 0 & c_d
\end{pmatrix}.
\end{equation}
In this scenario one loses some predictivity over the CKM mixing, since \eqref{eq:D15alteredCKM} leaves $ \lbrace \theta_{23}^q, \delta_{23}^q \rbrace$ unquantized --- they will become functions of the free operator couplings in a manner analogous to \eqref{eq:R23D15}.  On the other hand, the fine-tuning issue in $\tilde{\lambda}_{dl}$ is resolved, and \eqref{eq:D15alteredCKM} anyway better approximates global fits to the experimental CKM matrix than the original Cabibbo form we predicted.  Hence $\theta_{23}^q$ can be fit to the data, which then leads to more precise EFT predictions in the (unmeasured) leptoquark coupling of \eqref{eq:lamdlFPb}. 

\subsubsection*{Further Comments on the Appearance of Free Parameters in Effective Models}
We have seen that the simple equivalence evident in \eqref{eq:degenEV} can be important when building $\mathcal{L}_Y$ realizing family-symmetry breaking of the form $\mathcal{G_{F,L,Q}} \rightarrow \mathcal{G}_a   \cong \lbrace \hat{T}_a \rbrace$.  While this phenomenological ambiguity was discussed from a bottom-up perspective in \cite{Bernigaud:2019bfy} and multiple prior references from other authors, its consequences from a top-down model-building perspective have, to our knowledge, not been appreciated. We now see that, in the absence of a proper accounting of \eqref{eq:degenEV}, the implied RFS-invariant mass/coupling shapes are unnecessarily restrictive, possibly leading to the erroneous conclusion that fine-tunings of model parameters are required.  Upon considering the full implications of \eqref{eq:degenEV} on these shapes, these fine-tunings are resolved in favor of one-to-one mappings between model and physical parameters, albeit at the expense of the EFT's predictivity.  In short, the bottom-up mathematical ambiguity of \eqref{eq:degenEV} can consistently manifest itself as a top-down phenomenological ambiguity in a given model's IR mass and mixing spectrum.  

However, our $D_{15}$ analysis still leaves some questions unanswered.  For example, why was no tuning required in the neutrino or up quark mass matrices, where we also only attempted a map to the naive RFS-invariant mass matrices, but where Table \ref{tab:NADSresult} clearly indicates that free parameters can be introduced into these sectors as well?  While it is beyond our present scope to answer this question conclusively, we suspect that the answer lies in the group product rules at hand, which as a function of the group closure will (at least in the basis we consider) likely be driven by the CKM and PMNS structures entirely embedded in the up and neutrino sectors.  After all, $D_{15}$ is only armed with a handful of doublets from which we can form invariants according to \eqref{eq:fullayuka}, whereas a larger group that contains, e.g., triplet representations might allow a broader and more diverse set of invariants from which we can form the naive RFS-invariant shapes of \eqref{eq:D15genMass}.  This suspicion is at least consistent with the fact that, while studying $D_{15}$, we also attempted to build the final model presented in Table \ref{tab:NADSresult}, based on the same symmetry $\left(Z_{14} \times Z_2 \right) \rtimes Z_2$ that we used for the CKM prediction in Section \ref{sec:CompSM}.  There we again found that no tuning was required until we tried to model $m_l^{''\dagger} m^{''}_l$ and the subsequent leptoquark coupling $\lambda^{''}_{dl}$, where the need became apparent exactly as in $D_{15}$.  As conjectured, $\left(Z_{14} \times Z_2 \right) \rtimes Z_2$ also only has doublet and singlet irreducible representations.  Unfortunately though, we did not find a candidate symmetry in \cite{Bernigaud:2019bfy} that allows us to test to this hypothesis, and so it may be interesting to perform a similar group theory scan while allowing for larger finite groups to pass the self-imposed cuts.  On the other hand, the simultaneous introduction of the inherently 2D Cabibbo form of \eqref{eq:Cabibbo} alongside the inherently 3D $\mu-\tau$ symmetric form of \eqref{eq:mutauPMNS} into the scans may inevitably lead to similar results as those in \cite{Bernigaud:2019bfy}.  We will leave the resolution of these questions to future study --- our method, as demonstrated in Sections \ref{sec:THEORY}-\ref{sec:LEPTOQUARKS}, works regardless of their conclusions.

\section{Summary and Outlook}
\label{sec:CONCLUDE}
We have shown how to use the RFS of the Yukawa sector of an IR Lagrangian, i.e. one where electroweak- and family-symmetry breaking has already occurred, to systematically reconstruct a UV effective Lagrangian that respects SM gauge symmetries and non-Abelian flavour symmetries $\mathcal{G_F}$ which contain such RFS as subgroups.  Our method is thus complimentary to prior scans of family groups performed in order to identify phenomenologically viable $\mathcal{G_F}$ and symmetry-breaking patterns \emph{without} specifying concrete UV Lagrangians --- that is, we can use bottom-up, model-independent information to algorithmically construct top-down models with an explicit field and symmetry content.  We have shown four such examples, two where only SM fermionic mixing (CKM or PMNS matrices) is controlled, and two where SM mixing matrices and flavoured leptoquark couplings are structured with the RFS.  We thus provide `proof-in-principle' routes to EFT descriptions for the simplified models outlined in \cite{deMedeirosVarzielas:2019lgb,Bernigaud:2019bfy}.  Our study has also helped to clarify commentary in prior literature as regards the role of eigenvalue degeneracies in RFS generators and associated mixing ambiguities in top-down flavour models.

Furthermore, leptoquark extensions of the SM represent but one of many BSM scenarios with non-trivial flavour structure that can be studied within the RFS paradigm, which bypasses potentially unfalsifiable aspects of model building and offers a mechanism for identifying classes of simplified models and their phenomenological implications.  Our results in this paper indicate that analogous, model-independent RFS applications to (e.g.) multi-Higgs-doublet models (cf. \cite{deMedeirosVarzielas:2019dyu}) or softly-broken SUSY can also be readily `completed' if deemed necessary by a particular experimental signature, and can therefore be confidently studied in the meantime without reference to UV dynamics.  We leave these possible extensions to future work.

\section*{Acknowledgements}
JB's research was supported by the Deutsche Forschungsgemeinschaft (DFG, German Research Foundation) under grant 396021762 - TRR 257.
IdMV acknowledges
funding from Funda\c{c}\~{a}o para a Ci\^{e}ncia e a Tecnologia (FCT) through the
contract IF/00816/2015 and was supported in part by the National Science Center, Poland, through the HARMONIA project under contract UMO-2015/18/M/ST2/00518 (2016-2019), and by FCT through projects CFTP-FCT Unit 777 (UID/FIS/00777/2019), CERN/FIS-PAR/0004/2017 and PTDC/FIS-PAR/29436/2017 which are partially funded through POCTI (FEDER), COMPETE, QREN and EU.
JT gratefully acknowledges support from the Villum Fund, project number 00010102.

\begin{appendix}
\section{Details on Representation Theory}
\label{sec:APP}
In this Appendix we provide the relevant group product rules and associated Clebsch-Gordon coefficient structure for the finite groups employed in the sample models of the main paper, and in the bases in which they are built.\footnote{See e.g. \cite{Ishimori:2010au} for an exhaustive catalogue of finite group series and their properties.}  
\subsection{$A_4$}
\label{app:A4}
For two triplets generically parameterized by $x_i = \left(x_1, x_2, x_3 \right)$ and $y_i = \left( y_1, y_2, y_3\right)$, the product rule between them gives  
\begin{equation}
\bf{3} \times \bf{3} \sim \bf{1} + \bf{1^\prime} + \bf{1^{\prime \prime}} + \bf{3}_S + \bf{3}_A \,.
\end{equation}
In the basis of the generators given in \eqref{eq:GensAFFB}, the singlet Clebsch-Gordan structure is then given by 
\begin{align}
\nonumber
\bf{1} &\sim \left(x_1 y_1 + x_2 y_3 + x_3 y_2 \right)\,,\\
\nonumber
\bf{1^\prime} &\sim \left( x_3 y_3 + x_1 y_2 + x_2 y_1\right)\,,\\
\label{eq:singletproductA4}
\bf{1^{\prime \prime}} &\sim \left( x_2 y_2 + x_1 y_3 + x_3 y_1\right)\,,
\end{align}
whilst the symmetric ($S$) and anti-symmetric ($A$) triplet combinations are given by
\begin{align}
\nonumber
\bf{3}_S &\sim \frac{1}{3} \left(2x_1 y_1 - x_2 y_3 - x_3 y_2, 2x_3 y_3 - x_1 y_2 - x_2 y_1, 2x_2 y_2 - x_1 y_3 - x_3 y_1 \right)\,,\\
\label{eq:tripletproductA4}
\bf{3}_A &\sim \frac{1}{2} \left(x_2 y_3 - x_3 y_2, x_1 y_2 - x_2 y_1, x_1 y_3 - x_3 y_1 \right)\,.
\end{align}
Finally, the singlet product rules are found to be:
\begin{equation}
\bf{1^\prime} \times \bf{1^\prime} \sim  \bf{1^{\prime\prime}},\,\,\,\,\,\bf{1^\prime} \times \bf{1^{\prime\prime}} \sim  \bf{1},\,\,\,\,\,\bf{1^{\prime\prime}} \times \bf{1^{\prime\prime}} \sim  \bf{1^\prime} .
\end{equation}

\subsection{$\left(\mathbb{Z}_{14} \times \mathbb{Z}_2\right) \rtimes \mathbb{Z}_2$}
\label{app:complicated}
We have not found the associated product rules and Clebsch-Gordan factors for this group in the literature, and hence have derived them for ourselves.  As such, we provide a touch more information here than for other groups in this Appendix.

The relevant group information we require for $\left(\mathbb{Z}_{14} \times \mathbb{Z}_2\right) \rtimes \mathbb{Z}_2$ can be accessed via the {\tt GAP} package, using its identification number $\left[56, 7\right]$. However, {\tt GAP} is using a four-generator basis while the minimal generating set, that we will work with, only has three generators. Making this conversion, we find that the full list of irreducible representations and associated generators is given by
\begin{align}
\begin{split}
\mathbf{1_{++}} : \ \ \  &\mathbf{a} = 1, \ \ \ \mathbf{b} = 1, \ \ \ \mathbf{c} = 1 \,, \\
\mathbf{1_{+-}} : \ \ \  &\mathbf{a} = 1, \ \ \ \mathbf{b} = -1, \ \ \ \mathbf{c}=1 \,, \\
\mathbf{1_{-+}} : \ \ \  &\mathbf{a} = -1, \ \ \ \mathbf{b} = 1, \ \ \ \mathbf{c}=1 \,, \\
\mathbf{1_{--}} : \ \ \  &\mathbf{a} = -1, \ \ \ \mathbf{b} = -1, \ \ \ \mathbf{c}=1 \,, \\
 \mathbf{2_0} : \ \ \  &\mathbf{a} = \begin{pmatrix} 0 & 1 \\ 1 & 0 \end{pmatrix} \ \ \ \mathbf{b} = \begin{pmatrix} 1 & 0 \\ 0 & -1 \end{pmatrix}, \ \ \ \mathbf{c}=\begin{pmatrix} 1 & 0 \\ 0 & 1 \end{pmatrix} \,, \\
 \mathbf{2_{n++}} : \ \ \  &\mathbf{a} = \begin{pmatrix} 0 & 1 \\ 1 & 0 \end{pmatrix} \ \ \ \mathbf{b} = \begin{pmatrix} 1 & 0 \\ 0 & 1 \end{pmatrix}, \ \ \ \mathbf{c}=\begin{pmatrix} \omega_{7}^n & 0 \\ 0 & \omega_{7}^{-n} \end{pmatrix} \,, \\
 \mathbf{2_{n+-}} : \ \ \  &\mathbf{a} = \begin{pmatrix} 0 & 1 \\ 1 & 0 \end{pmatrix} \ \ \ \mathbf{b} = \begin{pmatrix} 1 & 0 \\ 0 & -1 \end{pmatrix}, \ \ \ \mathbf{c}=\begin{pmatrix} \omega_{7}^n & 0 \\ 0 & \omega_{7}^{-n} \end{pmatrix} \,, \\
 \mathbf{2_{n-+}} : \ \ \  &\mathbf{a} = \begin{pmatrix} 0 & 1 \\ 1 & 0 \end{pmatrix} \ \ \ \mathbf{b} = \begin{pmatrix} -1 & 0 \\ 0 & 1 \end{pmatrix}, \ \ \ \mathbf{c}=\begin{pmatrix} \omega_{7}^n & 0 \\ 0 & \omega_{7}^{-n} \end{pmatrix} \,, \\
\mathbf{2_{n--}} : \ \ \  &\mathbf{a} = \begin{pmatrix} 0 & 1 \\ 1 & 0 \end{pmatrix} \ \ \ \mathbf{b} = \begin{pmatrix} -1 & 0 \\ 0 & -1 \end{pmatrix}, \ \ \ \mathbf{c}=\begin{pmatrix} \omega_{7}^n & 0 \\ 0 & \omega_{7}^{-n} \end{pmatrix} \,, \\
\end{split}
\end{align}
where $n=1,2,3$ and $\omega_7 = e^{\frac{2i\pi}{14}}$.

As can be seen, representations are not always real. Taking the generic doublet to be $ \sim (x_1, x_2)$, conjugate representations can be expressed in terms of the original ones following 
\begin{equation}
\mathbf{\bar{2}_{k\rho\sigma}} \sim \begin{pmatrix} \bar{x}_2, \bar{x}_1 \end{pmatrix} \sim \mathbf{2_{k\sigma\rho}}\,,
\end{equation}
and similarly for $\mathbf{\bar{2}_0}$, while all the singlets are real.

The product rules for singlets can be obtained trivially by noting the action of the generators $\mathbf{a}$, $\mathbf{b}$, which is indicated by the first and second subscript respectively:
\begin{align}
\nonumber
\mathbf{1_{\pm \pm} \times 1_{\pm \pm}} &\sim \mathbf{1_{++}} \\
\nonumber
\mathbf{1_{\pm \mp} \times 1_{\pm \pm}} &\sim \mathbf{1_{+-}} \\
\nonumber
\mathbf{1_{\mp \pm} \times 1_{\pm \pm}} &\sim \mathbf{1_{-+}} \\
\mathbf{1_{\pm \mp} \times 1_{\mp \pm}} &\sim \mathbf{1_{--}}\,.
\end{align}

The full list of product rules can be obtained by explicitly checking the transformation properties. Assuming the first doublet is given by $(x_1, x_2)$ and the second one by $(y_1, y_2)$, we obtain the following product rules:
\begin{align}
\nonumber
\mathbf{2_0} \times \mathbf{2_0}  \,\,\, &\sim \,\,\, \left[ x_1 y_1 + x_2 y_2 \right]_{\mathbf{1_{++}}} + \left[ x_2 y_2 - x_1 y_1 \right]_{\mathbf{1_{-+}}} + \left[ x_1 y_2 + x_2 y_1 \right]_{\mathbf{1_{+-}}} + \left[ x_2 y_1 - x_1 y_2 \right]_{\mathbf{1_{--}}}\,, \\
\nonumber
\mathbf{2_{n++}} \times \mathbf{2_0}  \,\,\, &\sim \,\,\, \left[\begin{array}{cccc} x_1 y_1 \\ x_2 y_2 \end{array}\right]_{\mathbf{2_{n+-}}} + \left[\begin{array}{cccc} x_1 y_2 \\ x_2 y_1 \end{array}\right]_{\mathbf{2_{n-+}}} \,, \\
\nonumber
\mathbf{2_{n++}} \times \mathbf{2_{n++}}  \,\,\, &\sim \,\,\, \left[ x_1 y_2 + x_2 y_1 \right]_{\mathbf{1_{++}}} + \left[ x_2 y_1 - x_1 y_2 \right]_{\mathbf{1_{-+}}}  + \mathbf{2_{k ++}} \,, \\
\nonumber
\mathbf{2_{n++}} \times \mathbf{2_{n+-}}  \,\,\, &\sim \,\,\,   \left[\begin{array}{cccc} x_2 y_1 \\ x_1 y_2 \end{array}\right]_{\mathbf{2_0}} + \mathbf{2_{k +-}}
\,, \\
\nonumber
\mathbf{2_{n++}} \times \mathbf{2_{n-+}}  \,\,\, &\sim \,\,\,   \left[\begin{array}{cccc} x_1 y_2 \\ x_2 y_1 \end{array}\right]_{\mathbf{2_0}} + \mathbf{2_{k -+}}
 \,, \\
 \nonumber
\mathbf{2_{n++}} \times \mathbf{2_{n--}}  \,\,\, &\sim \,\,\,   \left[ x_2 y_1 + x_1 y_2 \right]_{\mathbf{1_{+-}}} + \left[ x_2 y_1 - x_1 y_2 \right]_{\mathbf{1_{--}}} + \mathbf{2_{k --}} \,, \\
\nonumber
\mathbf{2_{n+-}} \times \mathbf{2_0}  \,\,\, &\sim \,\,\, \left[\begin{array}{cccc} x_1 y_1 \\ x_2 y_2 \end{array}\right]_{\mathbf{2_{n++}}} + \left[\begin{array}{cccc} x_1 y_2 \\ x_2 y_1 \end{array}\right]_{\mathbf{2_{n--}}} \,, \\
\nonumber
\mathbf{2_{n+-}} \times \mathbf{2_{n+-}}  \,\,\, &\sim \,\,\,   \left[ x_2 y_1 + x_1 y_2 \right]_{\mathbf{1_{+-}}} + \left[ x_2 y_1 - x_1 y_2 \right]_{\mathbf{1_{--}}} + \mathbf{2_{k ++}}
\,, \\
\nonumber
\mathbf{2_{n+-}} \times \mathbf{2_{n-+}}  \,\,\, &\sim \,\,\,   \left[ x_2 y_1 + x_1 y_2 \right]_{\mathbf{1_{++}}} + \left[ x_2 y_1 - x_1 y_2 \right]_{\mathbf{1_{-+}}} + \mathbf{2_{k --}} \,, \\
\nonumber
\mathbf{2_{n+-}} \times \mathbf{2_{n--}}  \,\,\, &\sim \,\,\,   \left[\begin{array}{cccc} x_2 y_1 \\ x_1 y_2 \end{array}\right]_{\mathbf{2_0}} + \mathbf{2_{k -+}} \,, \\
\nonumber
\mathbf{2_{n-+}} \times \mathbf{2_0}  \,\,\, &\sim \,\,\, \left[\begin{array}{cccc} x_1 y_2 \\ x_2 y_1 \end{array}\right]_{\mathbf{2_{n++}}} + \left[\begin{array}{cccc} x_1 y_1 \\ x_2 y_2 \end{array}\right]_{\mathbf{2_{n--}}} \,, \\
\nonumber
\mathbf{2_{n-+}} \times \mathbf{2_{n-+}}  \,\,\, &\sim \,\,\,   \left[ x_2 y_1 + x_1 y_2 \right]_{\mathbf{1_{+-}}} + \left[ x_2 y_1 - x_1 y_2 \right]_{\mathbf{1_{--}}} + \mathbf{2_{k ++}} \,, \\
\nonumber
\mathbf{2_{n-+}} \times \mathbf{2_{n--}}  \,\,\, &\sim \,\,\,   \left[\begin{array}{cccc} x_1 y_2 \\ x_2 y_1 \end{array}\right]_{\mathbf{2_0}} + \mathbf{2_{k +-}} \,, \\
\nonumber
\mathbf{2_{n--}} \times \mathbf{2_0}  \,\,\, &\sim \,\,\, \left[\begin{array}{cccc} x_1 y_1 \\ x_2 y_2 \end{array}\right]_{\mathbf{2_{n-+}}} + \left[\begin{array}{cccc} x_1 y_2 \\ x_2 y_1 \end{array}\right]_{\mathbf{2_{n+-}}} \,, \\
\label{eq:CompDoubletRulesa}
\mathbf{2_{n--}} \times \mathbf{2_{n--}}  \,\,\, &\sim \,\,\,   \left[ x_2 y_1 + x_1 y_2 \right]_{\mathbf{1_{++}}} + \left[ x_2 y_1 - x_1 y_2 \right]_{\mathbf{1_{-+}}} + \mathbf{2_{k ++}}\,,
\end{align}
where 
\begin{equation}
\mathbf{2_{k \rho \sigma}} \sim 
\begin{cases} 
\left[\begin{array}{cccc} x_1 y_1 \\ x_2 y_2 \end{array}\right]_{\mathbf{2_{2\rho\sigma}}} & \text{for } n = 1 \\
\left[\begin{array}{cccc} x_2 y_2 \\ x_1 y_1 \end{array}\right]_{\mathbf{2_{3\sigma\rho}}} & \text{for } n = 2 \\
\left[\begin{array}{cccc} x_2 y_2 \\ x_1 y_1 \end{array}\right]_{\mathbf{2_{1\sigma\rho}}} & \text{for } n = 3
\end{cases}\,.
\end{equation}
While we expanded the above doublet product rules in \eqref{eq:CompDoubletRulesa} for simplicity and utility, we will express the other (less useful) doublet product rules in the following compacted formula:
\begin{equation}
\mathbf{2_{n \rho \sigma}}\times \mathbf{2_{n' \rho' \sigma'}} \sim \mathbf{2_X} + \mathbf{2_Y}\,,
\end{equation}
with $n\neq n'$ and
\begin{equation}
\begin{split}
&\mathbf{2_X} \sim 
\begin{cases} 
\left[\begin{array}{cccc} x_1 y_1 \\ x_2 y_2 \end{array}\right]_{\mathbf{2_{(n+n')(\rho \rho')(\sigma\sigma')}}} & \text{for } n+n' < 7-(n+n') \\
\left[\begin{array}{cccc} x_2 y_2 \\ x_1 y_1 \end{array}\right]_{\mathbf{2_{(7-n-n')(\sigma \sigma')(\rho\rho')}}} & \text{for } n+n' > 7-(n+n') 
\end{cases}\,, \\
&\mathbf{2_Y} \sim \begin{cases} 
\left[\begin{array}{cccc} x_1 y_2 \\ x_2 y_1 \end{array}\right]_{\mathbf{2_{(n-n')(\rho \sigma')(\sigma\rho')}}} & \text{for } \mathcal{F}_7 (n-n') < \mathcal{F}_7 (n'-n) \\
\left[\begin{array}{cccc} x_2 y_1 \\ x_1 y_2 \end{array}\right]_{\mathbf{2_{(n'-n)(\sigma\rho')(\rho \sigma')}}} & \text{for } \mathcal{F}_7 (n-n') > \mathcal{F}_7 (n'-n) 
\end{cases},
\end{split}
\end{equation}
where 
\begin{equation}
\begin{split}
\mathcal{F}_7 (k) = \begin{cases}
k & \text{for } k > 0 \\
7+k & \text{for } k < 0
\end{cases}\,.
\end{split}
\end{equation}

\subsection{$\Delta\left(96\right)$}
\label{app:D96}

We note that, as indicated above in Section \ref{sec:D96}, we follow the notation of \cite{Ding:2012xx} where the generators are identified by smallcase latin letters, but to help avoid confusion with our notation for coefficients (also smallcase latin letters), we refer to the generators of $\Delta(96)$ in bold.
Here we present some of the product rules for irreducible representations of $\Delta(96)$ in the basis we used in Section \ref{sec:D96}, which matches the basis in \cite{Ding:2012xx} where the element $\bf{a_{3_1}^2 c_{3_1} d_{3_1}}$ is diagonal. This basis has
\begin{equation}
\nonumber
\bf{a_{3_1}}=\frac{1}{3}\left(
\begin{array}{ccc}
 \omega  & 1+\sqrt{3} & \left(1-\sqrt{3}\right) \omega^2 \\
 1-\sqrt{3} & \omega^2 & \left(1+\sqrt{3}\right) \omega  \\
 \left(1+\sqrt{3}\right) \omega^2 & (1-\sqrt{3}) \omega & 1
\end{array}
\right) \,,
\end{equation}

\begin{equation}
\nonumber
\bf{b_{3_1}}=\frac{1}{3}\left(
\begin{array}{ccc}
 -1-\sqrt{3} & -\omega  & \left(\sqrt{3}-1\right) \omega^2 \\
 -\omega^2 & \sqrt{3}-1 & -\left(1+\sqrt{3}\right) \omega  \\
 \left(\sqrt{3}-1\right) \omega  & -\left(1+\sqrt{3}\right) \omega^2 & -1
\end{array}
\right)\,,
\end{equation}

\begin{equation}
\bf{c_{3_1}}=\frac{1}{3}\left(
\begin{array}{ccc}
 1 & \left(1-\sqrt{3}\right) \omega^2 & \left(1+\sqrt{3}\right) \omega  \\
 \left(1+\sqrt{3}\right) \omega  & 1 & \left(1-\sqrt{3}\right) \omega^2 \\
 \left(1-\sqrt{3}\right) \omega^2 & \left(1+\sqrt{3}\right) \omega  & 1
\end{array}
\right)\,,
\end{equation}
and $\bf{d_{3_1} = a_{3_1}^{-1} c_{3_1} a_{3_1}}$. In this basis, the group rules we have used in our models are all the rules for $\mathbf{3_1 \times \bar{3}_1 } \sim \mathbf{1 + 2 + 6}$,  $\mathbf{3_1 \times \bar{3}'_1 } \sim \mathbf{1' + 2 + 6}$, and only the rules for the trivial singlet built from
$\mathbf{(1' \times 1')} \sim \mathbf{1}$, $\mathbf{(2 \times 2)_1}$ and $\mathbf{(6 \times 6)_1}$:
\begin{align}
\nonumber
\mathbf{(3_1 \times \bar{3}_1)_1} \, &\sim \, x _1  y _1+x _2  y _2+x _3  y _3 \,, \\
\nonumber
\mathbf{(3_1 \times \bar{3}_1)_2 } \, &\sim \, \left(
\begin{array}{c}
x_1  y _3+x _2  y _1+x _3  y _2 \\
 \omega\left( x _1  y _2+x _2  y _3+x _3  y _1\right)
\end{array}
\right) \,, \\
\nonumber
\mathbf{(3_1 \times \bar{3}_1)_6 } \, &\sim \, \left(
\begin{array}{c}
x _1  y _3  +\omega x _2  y _1 +\omega^2 x _3  y _2 \\
 \omega x _1  y _2+\omega^2 x_2  y _3+x _3  y _1   \\
 \omega^2 x_1  y _1+x _2  y _2  +\omega x _3  y _3  \\
 x _1  y _3  +\omega^2 x_2  y _1+ \omega x _3  y _2  \\
 \omega^2 x_1  y _2+\omega x _2  y _3 +x _3  y _1   \\
 \omega x _1  y _1 +x _2  y _2  +\omega^2 x _3  y _3
\end{array}
\right) \,, \\
\nonumber
\mathbf{(3_1 \times \bar{3}_1)_1'} \, &\sim \,  x _1  y _1+ x _2  y _2+ x _3  y _3 \,, \\
\nonumber
\mathbf{(3_1 \times \bar{3}'_1)_2} \, &\sim \, \left(
\begin{array}{c}
 x_1  y _3+ x _2  y _1+ x _3  y _2 \\
-  \omega\left(  x _1  y _2+ x _2  y _3+ x _3  y _1\right)
\end{array}
\right) \,,\\
\nonumber
\mathbf{(3_1 \times \bar{3}'_1)_6 } \, &\sim \, \left(
\begin{array}{c}
 x _1  y _3+\omega x _2  y _1+\omega^2 x _3  y _2 \\
 \omega x _1  y _2 +\omega^2 x _2  y _3+ x _3  y _1  \\
 \omega^2 x _1  y _1+ x _2  y _2+\omega x _3  y _3   \\
 - x _1  y _3 -\omega^2 x _2  y _1-\omega x _3  y _2  \\
 -\omega^2 x _1  y _2-\omega x _2  y _3 - x _3  y _1   \\
 -\omega x _1  y _1 - x _2  y _2 -\omega^2 x_3  y _3
\end{array}
\right) \,, \\
\nonumber
\mathbf{(2 \times 2)_1 } \, &\sim \,  x _1  y _2+ x _2  y _1 \,,\\
\mathbf{(6 \times 6)_1} \, &\sim \,  x _1  y _5+ x _2  y _4+ x _3  y _6+ x _4  y _2+ x _5  y _1+ x _6  y_3 \, .
\end{align}

\subsection{$D_{15}$}
\label{app:D15}

Following \cite{Ishimori:2010au}, $D_N$ has $2 + (N-1)/2$ representations when $N$ is odd: 2 singlets $\mathbf{1_+}$ and $\mathbf{1_-}$ ($\mathbf{1_+}$ is the trivial singlet) and $(N-1)/2$ doublets labeled $\mathbf{2}_k$. In our case we therefore have 7 doublet representations,
within which our group generators take the form
\begin{align*}
\mathbf{b} = \begin{pmatrix}
1 & 0\\
0 & -1
\end{pmatrix}, \ \ \ \ 
\mathbf{a}=\begin{pmatrix}
\cos \frac{2k\pi}{15} & -\sin \frac{2k\pi}{15}\\
\sin \frac{2k\pi}{15} & \cos \frac{2k\pi}{15}
\end{pmatrix}.
\end{align*}
On the other hand, for the singlet representations we have that
\begin{equation}
\begin{split}
\mathbf{1_+} &: \ \ \ \ \mathbf{a} = 1 \ \ \ \text{and} \ \ \ \ \mathbf{b}=1,\\
\mathbf{1_-} &: \ \ \ \ \mathbf{a} = 1 \ \ \ \text{and} \ \ \ \ \mathbf{b}= -1 .
\end{split}
\end{equation}
One can work out the kronecker products for these different representations. In our case, the relevant ones for contracting two doublets will be given by
\begin{align}
\nonumber
 \left( \begin{pmatrix} x_1 \\ x_2 \end{pmatrix} _{\mathbf{2_k}} \times \begin{pmatrix}y_1 \\ y_2 \end{pmatrix}_{\mathbf{2_k}}  \right)_{\mathbf{1}} &\sim x_1 y_1 + x_2 y_2, \\
 \left( \begin{pmatrix} x_1 \\ x_2 \end{pmatrix} _{\mathbf{2_k}} \times \begin{pmatrix}y_1 \\ y_2 \end{pmatrix}_{\mathbf{2_k}}  \right)_{\mathbf{1'}} &\sim x_2 y_1 - x_1 y_2,
 \end{align}
 while two non-trivial singlets contract to a trivial singlet
 \begin{align}
\mathbf{1'} \times \mathbf{1'} &\sim \mathbf{1} .
\end{align}

\section{Details on Flavon VEV Condition}
\label{sec:APPB}
In this Appendix we present a derivation of the core RFS-preserving condition on the flavon VEV we impose in \eqref{eq:vevconstrain}.  While this condition has been known since as early as \cite{Lam:2007qc}, we now show our own approach for clarity and completeness.

We start with the generic $\mathcal{G}$-invariant term that will lead to a mass matrix,
\begin{equation}
\mathcal{L} = C^{A a \alpha} F_{A}^{\rho_F}  f_{a}^{\rho_f} \phi_{\alpha}^{\rho_\phi} \,,
\end{equation}
where $F_{A}^{\rho_F}$ and $f_{a}^{\rho_f}$ are respectively $SU(2)_L$ doublets and singlets with flavour representations $\rho_F$ and $\rho_f$. $\phi_{\alpha}^{\rho_\phi}$ is the flavon field with representation $\rho_\phi$. Finally, $C^{A a \alpha} = C_{\rho_F \times \rho_f \times \rho_\phi \rightarrow 1}^{A a \alpha}$ stands for the Clebsch-Gordan matrix that pins down the product representation to a specific flavour singlet.

Let's act on the term with an element $g \in \mathcal{G}$, which leaves the singlet invariant:
\begin{gather}
g \left(C^{A a \alpha} F_{A}^{\rho_F}  f_{a}^{\rho_f} \phi_{\alpha}^{\rho_\phi}\right) = C^{C c \gamma} F_{C}^{\rho_F}  f_{c}^{\rho_f} \phi_{\gamma}^{\rho_\phi}\,,
\\
C^{A a \alpha}\, T_{AB}^{\rho_F}(g) \, F_{B}^{\rho_F}\, T_{ab}^{\rho_f}(g)\,  f_{b}^{\rho_f} \, T_{\alpha_\beta}^{\rho_\phi}(g)\, \phi_{\beta}^{\rho_\phi} = C^{C c \gamma} F_{C}^{\rho_F}  f_{c}^{\rho_f} \phi_{\gamma}^{\rho_\phi}\,.
\end{gather}
In the broken phase, where $ \mathcal{G} \rightarrow \mathcal{H}$ (where $\mathcal{H}\subset  \mathcal{G}$) by $\langle \phi_\alpha \rangle$ with the condition 
\begin{equation}
T_{\alpha \beta}^{\rho_\phi} \langle \phi_\beta \rangle = \langle \phi_\alpha \rangle\,,
\label{Eq:vev_condition_1}
\end{equation}
the mass matrix $m^{Aa} = C^{Aa\alpha}\langle \phi_\alpha \rangle$ exhibits an invariance under generic elements $h\in \mathcal{H}$,
\begin{equation}
F_{B}^{\rho_F}\, T_{AB}^{\rho_F}(h)\, m^{Aa}\, T_{ab}^{\rho_f}(h)\,  f_{b}^{\rho_f} = F_{C}^{\rho_F}\, m^{Cc}  \,  f_{c}^{\rho_f} \,.
\end{equation}
Moreover, when we consider the broken element transformations $T(g')$ where $g' \notin \mathcal{H}$, the VEV instead transforms as $T_{\alpha \beta} \langle \phi_\beta \rangle = \langle \phi'_\alpha \rangle \Rightarrow m\rightarrow m'$. This leads to the following equality:
\begin{equation}
F_{B}^{\rho_F}\, T_{AB}^{\rho_F}(g')\, m'^{Aa}\, T_{ab}^{\rho_f}(g')\,  f_{b}^{\rho_f} = F_{C}^{\rho_F}\, m^{Cc}  \,  f_{c}^{\rho_f} \,,
\end{equation}
which explicitly shows the non-invariance of the mass matrix under $g'$.

The invariance relation is also easily extended to the combination $m m^\dagger$, which is the more general framework that we consider in the paper. Starting from the relation
\begin{equation}
T_{AB}^{\rho_F}(h)\, m^{Aa}\, T_{ab}^{\rho_f}(h) =  m^{Bb}   \,,
\end{equation}
we end up with
\begin{equation}
T_{AB}^{\rho_F}(h)\, m^{Ba} (m^\star)^{aC}\, T_{CD}^{*\rho_F}(h) =  m^{Ab}  (m^*)^{bD} \,.
\end{equation}
In that case $T_{AB}^{\rho_F}(h)\rightarrow T_{AB}^{\rho_F}(h) e^{i\theta}$ leaves the condition invariant; therefore the VEV preserving combination $mm^\dagger$ condition is simply given by
\begin{equation}
T_{\alpha \beta}^{\rho_\phi}\, e^{i\theta} \langle \phi_\beta \rangle = \langle \phi_\alpha \rangle\,.
\label{Eq:vev_condition_2}
\end{equation}
Here one then clearly sees the origin for the condition $\hat{e}_a^\star \cdot \hat{e}_a \overset{!}{=} 1$ in \eqref{eq:vevconstrain}.
Finally, the corresponding constraint for Majorana mass terms can be easily derived from the same procedure, and in that case one finds that no additional phases are present.
\end{appendix}

\end{document}